\documentclass{aa}
\usepackage{color}
\usepackage{graphicx}
\usepackage[fleqn]{amsmath}	
\usepackage{amssymb}	
\usepackage{mathtools}
\usepackage{txfonts}
\usepackage{natbib,twoopt}

\usepackage{float}
\usepackage[breaklinks=true]{hyperref}
\hypersetup{
  colorlinks=true,   
  citecolor=blue,    
  urlcolor=blue,     
  linkcolor=blue,     
}
\bibpunct{(}{)}{;}{a}{}{,} 

\newcommand{\lo}{L_{\rm UV}}
\newcommand{\lx}{L_{\rm X}}
\newcommand{\fo}{F_{\rm UV}}
\newcommand{\fx}{F_{\rm X}}
\newcommand{\om}{\Omega_{\rm M}}
\newcommand{\ol}{\Omega_\Lambda}

\begin{document} 

   \title{Are quasars reliable standard candles?}
   
   \author{
   E. Lusso\inst{1,2}\thanks{\email{elisabeta.lusso@unifi.it}}, G. Risaliti\inst{1,2}, E. Nardini\inst{2}
          }

\institute{
$^{1}$ Dipartimento di Fisica e Astronomia, Universit\`{a} di Firenze, via G. Sansone 1, 50019 Sesto Fiorentino, Firenze, Italy\\
$^{2}$ INAF-Osservatorio Astrofisico di Arcetri, Largo E. Fermi 5, 50125, Firenze, Italy
}

   \date{\today}

 
  \abstract{In this paper we address the question whether the non-linear relation between the X-ray and UV emission of quasars can be used to derive their distances. In previous works of our group, we demonstrated that such a relation does not show any redshift evolution in its slope. 
  The derived distances are in agreement with the standard flat $\Lambda$CDM model up to $z$\,$\sim$\,1.5, but they show significant deviations at higher redshifts. Yet, several authors suggested that this discrepancy is due to inconsistencies between the low- and high-redshift sources within the parent sample, or to a redshift evolution of the relation. Here, we discuss these issues through a quantitative comparison with supernova-derived distances in the common redshift range, complemented by simulations showing that all the claimed inconsistencies would naturally arise from any limitation of the cosmological model adopted for the data analysis, that is, from our ignorance of the {\it true} cosmology. We argue that the reliability of the method can only be based on a cosmology-independent evaluation of the hypothesis of non-evolution of the X-ray to UV relation at $z$\,$>$\,1.5, subsequent to a careful check of the sample selection and of the flux measurements for possible redshift-dependent systematic effects. Since we do not conceive any physical reason for a sudden change of the normalization of the relation at $z$\,$>$\,1.5, and we can exclude any severe systematic effect in the data selection and flux measurements, we conclude that the application of the X-ray to UV relation to cosmology is well motivated. To further strengthen this point, we need to achieve a better understanding of the physical process behind the observed relation and/or an independent observational proof possibly confirming the discrepancy with $\Lambda$CDM found with quasars, such as future supernova measurements at $z$\,$\sim$\,2 or higher.}


   \maketitle
%

\section{Introduction}
The cosmological concordance model, often called Lambda-CDM, or $\Lambda$CDM, where CDM stands for `cold dark matter' and $\Lambda$ represents the cosmological constant, is the best description of the current observations. The accelerated expansion of the Universe \citep[see also \citealt{Weinberg2013} for a review]{riess1998,perlmutter1999}, the cosmic microwave background (CMB; e.g., \citealt{Bennett1996}), and the measurement of baryon acoustic oscillations (BAO; see \citealt{Eisenstein2005} for a review) are key predictions of $\Lambda$CDM, which led to its general consensus amongst the scientific community. Nonetheless, discrepancies with $\Lambda$CDM have also been observed, the  prevailing ones being the `$\sigma_8$ tension' and the `Hubble tension' (see \citealt{Abdalla2022} for a review on cosmological tensions and anomalies). 
Most recently, cosmological results from the first year of galaxy, quasar, and Lyman-$\alpha$ forest BAO measurements from the Dark Energy Spectroscopic Instrument (DESI Data Release 1) have shown a preference for $w_0$\,$>$\,$-1$ and $w_a$\,$<$\,0 when a time-varying energy density equation of state $w(z)=w_0+w_a\,z/(1+z)$ is considered, which departs from $\Lambda$CDM at about the $3\sigma$ statistical level, depending on the specific data set used \citep{desivi2025}. 
The same overabundance of very high redshift galaxies that are being discovered by the James Webb Space Telescope (e.g.,  \citealt{Finkelstein2023,Harikane2023}) poses several challenges to the timescale of structure formation in a $\Lambda$CDM Universe \citep[e.g.,][]{BK2023,melia2023}.  
In this landscape, our analysis of the distance--redshift relation (also referred to as the Hubble-Lemaître diagram) of luminous, unobscured active galactic nuclei (AGN, where quasars represent the most powerful members of this population) has also revealed a deviation from the predictions of the flat $\Lambda$CDM model, which emerges at high redshift with a statistical significance of 3--4$\sigma$ \citep[][see also \citealt{bargiacchi2022}]{rl19,lusso19,lusso2020}.
While complementary to the traditional resort to supernovae to estimate the cosmological parameters, quasars extend the distance--redshift diagram to a redshift range inaccessible to supernovae ($z$\,$=$\,2--7.5). All the above-mentioned findings, if confirmed, would be important indications for the need of new physics beyond $\Lambda$CDM. 

The determination of the distance versus redshift relation for quasars, and thus of the cosmological model parameters, is obtained from the empirical correlation between the X-ray and UV emission observed in AGN \citep{avnitananbaum79,zamorani81,avnitananbaum82,avnitananbaum86}. Such a relation is non-linear, and has been observed over almost 5 orders of magnitude in UV luminosity with a dispersion of just $\sim$\,0.2 dex, implying that there must be a ubiquitous physical mechanism that regulates the energy transfer from the accretion disc to the X-ray emitting corona \citep[e.g.,][and references therein]{lr17}. Several models have been proposed to explain the physics behind the X-ray to UV relation, involving, for example, reprocessing of radiation from a non-thermal electron-positron pair cascade \citep{1982ApJ...258..321S,1984MNRAS.209..175S,1983MNRAS.205..593G,1990ApJ...363L...1Z}, a two-phase accretion disc model where the entire gravitational power is dissipated via buoyancy and reconnection of magnetic fields in a uniform hot plasma in close vicinity of the cold opaque disc \citep{1991ApJ...380L..51H,1993ApJ...413..507H,1994ApJ...436..599S,1998MNRAS.299L..15D}, or a viscosity-heated corona \citep{2000A&A...361..175M,2002ApJ...575..117L}. Magnetic field turbulence has also been recognised not only as a supplementary heating process in the formation of the corona itself (e.g., \citealt{1979ApJ...229..318G,mf2001,2002ApJ...572L.173L}), but also as an efficient means for the transport of the disc angular momentum (e.g., \citealt{2003ARA&A..41..555B}).

Although a comprehensive model is still missing, and the dependence of the coronal properties on black hole mass ($M_{\rm BH}$) and accretion rate ($\dot{M}$) remains an open question, there is little doubt that the X-ray to UV relation is {\it inherent} to the accretion process itself, and that a tight physical connection exists between these two bands in the spectral energy distribution (SED) of AGN. We note that relations of some kind are known between various pairs of wavebands, virtually including the AGN emission in any part of the SED from radio to gamma-rays, yet these relations involve a combination of different processes and physical scales, and do not automatically hold any cosmological value. Indeed, the use for cosmological purposes of any relation between the emission in two given bands must obey several conditions, besides the non-linearity of its slope. First and foremost, the dispersion should be small, implying a nearly {\it consequential} connection (albeit not necessarily known) between the two bands. Moreover, while the correlation has to be observed across a wide redshift range and for numerous sources, its key parameters should not show any evolution with redshift. Here we aim to demonstrate that the correlation between X-ray and UV emission in AGN is rooted in black-hole accretion physics and boasts all the aforementioned features, therefore being a valid cosmological tool irrespective of its largely empirical nature.

Given the apparent tension with $\Lambda$CDM of the distance--redshift relation of high-redshift quasars, it is mandatory to investigate whether hidden systematics, uncertainties in the data, or possible biases may account for such a discrepancy. Criticism was in fact raised by some authors regarding the reliability of this result. \cite{KR2020,KR2021,KR2022} suggested that the disagreement of the quasar distance--redshift relation with respect to the prediction of the flat $\Lambda$CDM model at high redshift can be due to the heterogeneity of the quasar sample considered in the analysis of \citet{lusso2020}, and specifically to some issues with the high-redshift subsamples. Their analysis was performed assuming different cosmological models, indicating that the values of the parameters involved in the determination of quasar distances ($\gamma$ and $\beta$, see Section~\ref{sect:method}) are dependent upon the cosmological model assumed. This, together with the possible evolution with redshift of the same two parameters, would prevent the standardisation of quasars, hence the applicability of the distances derived in this fashion to cosmological studies \citep[see also][]{singal2022,petrosian2022}. In this paper, we address all the above issues through a quantitative comparison with supernova-derived distances in the common redshift range, complemented by simulations that prove that all the claimed shortcomings are not intrinsic to the data, but naturally arise when the cosmological model assumed in the analysis is incorrect. In other words, the purported non-standardisability of high-redshift quasars is a direct consequence of our ignorance of the {\it true} cosmological model.

The manuscript is structured as follows. In Section~\ref{sect:qsosample} we briefly describe our latest quasar sample used for cosmological analyses, whilst our method is summarised in Section~\ref{sect:method}. Section~\ref{sect:qso-sn comparison} presents a quantitative comparison of the quasar distances obtained through our technique with the ones of type Ia supernovae in the common redshift range, and in Section~\ref{sect:gammaevolution} we investigate any possible redshift evolution of the relation used to standardise quasars.
Section~\ref{sect:Simulations} includes a set of simulations to demonstrate the intrinsic degeneracy between the determination of the cosmological parameters assuming different input values for $\om$ and $\ol$ and a redshift evolution of the X-ray to UV relation. In Section~\ref{sect:On the intrinsic luminosity evolution} we discuss the issue of luminosity evolution.
In Section~\ref{sect:General guidelines} we provide general guidelines that should be followed to use quasars as standardisable candles for cosmology. 
Our conclusions are drawn in Section~\ref{sect:conclusion}.


\section{The quasar sample}
\label{sect:qsosample}
The quasar sample published by \citet[][hereafter L20]{lusso2020} contains 2421 sources over the redshift range 0.009\,$\leq$\,$z$\,$\leq$\,7.541. The interested reader should refer to Section 7 in L20 for a detailed discussion of the selection criteria. The sample was intended to be highly uniform, as radio-bright, broad-absorption-line, and significantly absorbed quasars at UV and/or X-ray energies were neglected. 
Nonetheless, some residual contamination from the host galaxy emission in the optical can still be present, especially for the low-redshift ($z$\,$\lesssim$\,0.5) and low-luminosity ($\lesssim$\,10$^{45}$ erg s$^{-1}$) AGN, for which the contrast between the nuclear and the host-galaxy emission is limited. This additional contribution biases the determination of the continuum emission from the disc, which, for these sources, has to be extrapolated to the rest-frame 2500 \AA. 
Therefore, unless ultraviolet data are available for the low-redshift AGN (i.e., covering the rest-frame 2500 \AA, see Section 2.7 in L20), AGN at $z$\,$<$\,0.7 should be conservatively excluded from the sample that is used for the calibration of quasars with supernovae, and also from the cosmological analysis. 
In this paper, we thus use 2036 quasar measurements, i.e., the 2023 quasars at $z$\,$>$\,0.7 plus the 13 AGN at very low redshift
(0.009\,$<$\,$z$\,$<$\,0.087) with UV data from the International Ultraviolet Explorer (IUE) in the Mikulski Archive for Space Telescopes (MAST), for which $\lo$ could be determined directly from the spectra (see Section 2.7 in L20).

\section{The method: computation of luminosity distances}
\label{sect:method}
Whilst the methodology we adopt to derive quasar distances has been presented in depth in a series of works \citep[e.g.][]{rl15,rl19,lusso2020,moresco2022}, we briefly summarise the main points below.
The method is based on the hypothesis of a universal, redshift-independent, non-linear relation between the X-ray and UV luminosities of quasars:
\begin{equation}
\label{lxlo}
    \log(\lx) = \gamma \log(L_{\rm UV}) + \beta,
\end{equation}
with $\gamma$\,$\simeq$\,0.6 and $\beta$\,$\simeq$\,8.0. Note that the exact values, at this point, are assumed to be an inherent attribute of the accretion process, and are therefore cosmology-independent. We will justify this assumption at length in the following. The X-ray to UV relation has been observed since the '80s \citep{avnitananbaum79,avnitananbaum82,avnitananbaum86} and widely studied ever since by a number of independent groups and by using different sample selections. In this paper we will use the data in L20, where $\lx$ and $\lo$ are the monochromatic luminosities at the rest-frame 2~keV and 2500~\AA, respectively, as derived from X-ray and UV photometric measurements. We extensively discussed the choice of the optimal indicators of the X-ray (coronal) and UV (accretion disc) emission in \citet[][see also \citealt{jin2024}]{signorini2023}. While we refer to that work for more details, we have shown that the physical quantities that are more tightly linked to one another are the soft X-ray flux at $\sim$\,1 keV and the ionising UV flux blueward of the Lyman limit, and yet the usual monochromatic fluxes at 2 keV and 2500 \AA\ estimated from photometric data provide an almost as tight X-ray to UV relation, which can thus be used to derive reliable quasar distances \citep{signorini2023}. 

If the relation in equation (\ref{lxlo}) holds at all redshifts (we will discuss this key point further in Section~\ref{sect:gammaevolution}), then it is straightforward to derive the relation between X-ray and UV fluxes:
\begin{equation}
\label{fxfo}
    \log(\fx) = \gamma \log(\fo) + \hat\beta,
\end{equation}
where 
\begin{equation}
\label{betahat}
\hat\beta(z) = \beta-(1-\gamma)\log(4\pi)-(2-2\gamma) \log D_L(z),
\end{equation}
and $D_L$ is the luminosity distance.\footnote{Note that the quantities $\beta$ and $\hat\beta(z)$ are the same as $\beta_{L}(z)$ and $\beta_{F}(z)$, respectively, in the formalism of \citet{petrosian2022}.}
It is therefore possible to fit the data (i.e., $\fx$, $\fo$, $z$) for each quasar, leaving the cosmological parameters of the model describing the $D_L(z)$ expression and the variables $\gamma$ and $\beta$ of the $\lx-\lo$ relation free to vary.

We reiterate that for quasars to be standardisable candles, and thus for this technique to work, the correlation parameters $\gamma$ and $\beta$ must be \emph{intrinsically} cosmology-independent. To validate (or to confute) the reliability of this method, the key question is whether this hypothesis holds. The rejection of this premise immediately leads to the {\it circularity} argument raised by \citet{petrosian2022}. In fact, by reading equation~(\ref{lxlo}) at face value, both $\gamma$ and $\beta$ are dependent upon the choice of the cosmological model. Different assumptions clearly result in slightly different values of the {\it observed} correlation parameters when analysing any real data set. This is not in contradiction with the underlying hypothesis, as the following analysis, entirely based on equations~(\ref{fxfo}) and (\ref{betahat}), is devised to prove. 

It is also worth noting that for any {\it prospective} standard candle, a direct, observational validation is impossible in the redshift range where no other {\it established} standard candle is present. For instance, one can observationally approve the distance measurements from type Ia supernovae only if these are found in galaxies close enough to have Cepheid-based distances available (up to 40 Mpc; \citealt{riess2016}).  
For supernovae further away, the reliability of the distance measurements is based on two conditions: a deep physical understanding of the process leading to the distance estimates (i.e., the physical grounds of the Phillips relation; \citealt{phillips1993}), and the absence of observational and selection biases in the sample. 
In the case of type Ia supernovae, there is now a general consensus on both the above points, so their use as distance indicators is well-established and the Hubble diagram of type Ia supernovae represents one of the pillars of observational cosmology. 
The situation is certainly different for quasars. 
In particular, there is not a satisfactory physical explanation of the X-ray to UV relation yet, which should necessarily involve the process of energy transfer from the accretion disc to the X-ray emitting corona to link the values of $\gamma$ and $\beta$ to the workings of accretion physics. 
To assess the reliability of the distance estimates based on the X-ray to UV relation, we need to: (1) rule out any possible redshift-dependent selection effect in the sample, including any evolution in the spectroscopic properties of the selected sources; (2) check for any possible systematic trend in the common redshift range with supernovae; and, based on the previous  points, (3) define a method to use quasars as standardisable candles in combination with other cosmological probes.

The sample selection (first point above) has been already discussed in great detail in our previous works \citep{lr16,rl19,lusso2020}, while the analysis of the spectroscopic properties of the sample as a function of redshift, black-hole mass, and luminosity was presented in \citet{trefoloni2024}.
\citet{sacchi2022} specifically focused on the individual spectroscopic analysis of the $z$\,$>$\,2.5 quasars. 
All these studies found that, when typical blue quasars are selected, the X-ray and UV continuum emission and the overall spectral properties are remarkably similar regardless of redshift. Moreover, when the X-ray and UV continuum luminosities are determined from a one-by-one analysis, the dispersion in the $\lx-\lo$ relation narrows down significantly, further hinting at an underlying {\it causal} connection.

Here, we will concentrate on a quantitative discussion of the second point, and on the general guidelines for the third point. 
Our main goals are to uphold the use of quasars as standardisable candles in the light of the criticism recently advanced in some papers; and to lay down a set of conservative rules that should be followed for a safe use of quasars in cosmology, avoiding possible biases or, even worse, incorrect results. 
Specifically, the essential issues we want to address are the following: 
\begin{itemize}
    \item The Hubble diagram of quasars, if not analysed in combination with supernovae, prefers a value of the present total matter density parameter of $\om$\,$>$\,0.6, that is, significantly higher than the concordance value of $\om$\,$\simeq$\,0.3 obtained through supernovae and BAO \citep[see, e.g.,][]{bargiacchi2022}. This could suggest a disagreement between the Hubble diagrams of quasars and supernovae \citep{KR2020}. 
    
    \item When the X-ray to UV relation is studied in narrow redshift intervals to factor out any cosmological assumption, the finite width of the bins can introduce biases in the determination of the correlation parameters \citep{petrosian2022}. 
    
    \item A discrepancy on slope and intercept values of the X-ray to UV relation emerges when the Hubble diagram of quasars is studied assuming different cosmological models (including $\Lambda$CDM and several of its extensions). This is also observed when the L20 sample is split in two subsamples at low and high redshift (with a separation value of $z$\,$\simeq$\,1.5; see \citealt{KR2020,KR2021}). These results have been ascribed to the non-standardisability of the quasar sample, and may stem from a redshift evolution of the relation. Some groups have proposed a correction for the possible evolution of the luminosities in the cosmological fits \citep[e.g.,][]{singal2022,dainotti2022,dainotti2023,wang2022,wang2024}.  
\end{itemize}
While we will thoroughly discuss all these concerns in the following sections, it is also worth noting that other groups have instead tried to improve the precision on the determination of the cosmological parameters resorting to our technique \citep[e.g.,][]{melia2019,colgain2024arXiv}. Indeed, despite the limitation represented by their large dispersion in the Hubble diagram ($\sim$\,1.4 dex) with respect to supernovae ($<$\,0.1 dex), quasars are complementary to the latter as cosmological probes: they come in much greater numbers and they offer a unique opportunity to explore the very first billion years of the Universe.

\section{Comparison of the Hubble diagrams of quasars and type Ia supernovae}
\label{sect:qso-sn comparison}

A quantitative comparison of quasars and supernovae as distance indicators is not straightforward. 
If one derives the luminosity distance $D_L$ of quasars from equations (\ref{fxfo}) and (\ref{betahat}), the values of $D_L$ will depend on the parameters $\gamma$ and $\beta$ (see Section~\ref{Cosmological analysis: the likelihood}).  
The parameter $\gamma$ controls the shape of the Hubble diagram, while the value of $\beta$ is degenerate with the absolute scale of $D_L$, which can be obtained only through the cross-calibration with other standard candles in the common redshift range.
This is why quasars should always be cross-calibrated with supernovae to extend the distance ladder in a similar way to what is done for supernovae with Cepheids. 

Crucially, the intrinsic value of the slope $\gamma$ can be estimated independently from any cosmological fit by analysing the X-ray to UV relation in small redshift bins. We adopted this procedure in several previous papers \citep[e.g.,][]{rl15,rl19,lusso2020} to test whether any evolution of the slope of the relation with redshift can be detected. We performed this test by analysing the $\fx-\fo$ relation in narrow redshift intervals, such that the differences in distance are smaller than the dispersion of the relation within each bin.  
We will demonstrate that this condition is met in Section~\ref{subsect:gamma-z}; here we just want to analyse the consistency between the Hubble diagram of supernovae and the one of quasars in the redshift range where both populations are present. We can perform this test over the redshift interval 0.7--1.6, which contains 1157 quasars, i.e., $\sim$\,50\% of the whole sample of 2421 objects in L20. 

\begin{figure}[h!]
\centering
\includegraphics[width=\linewidth,clip]{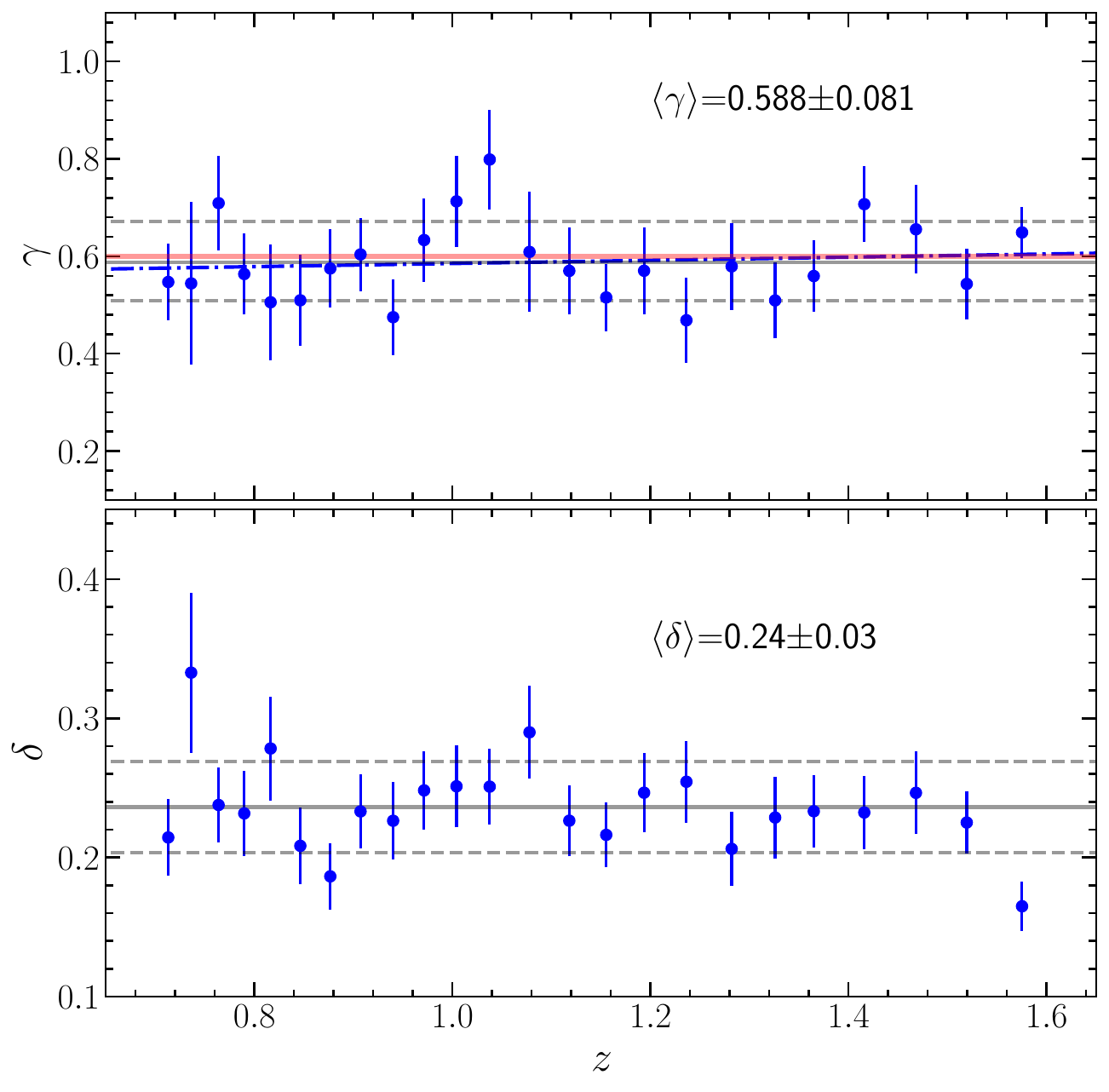}
\caption{Best-fit slope and intrinsic dispersion from the regression analysis of the $\fx-\fo$ relation in narrow redshift intervals from $z$\,$=$\,0.7 to $z$\,$=$\,1.6. The red solid line represents $\gamma$\,$=$\,0.6. A linear fit of $\gamma$ as a function of redshift in the form $\gamma=m\,z+q$ gives $m=0.032\pm0.067$ and $q=0.553\pm0.075$ (dot-dashed blue line).}
\label{fig:qsolowz}
\end{figure}

We have fitted the $\fx-\fo$ relation in narrow redshift bins, and the best-fit slope and intrinsic dispersion from the regression analysis are plotted in Figure~\ref{fig:qsolowz}. As evident from this figure, there is no redshift evolution, where this is intended as a systematic trend and not just as an empirical scatter of the slope: a linear fit as a function of redshift, of the form $\gamma=m\,z+q$, returns $m=0.032\pm0.067$ and $q=0.553\pm0.075$. The statistical analysis of the quasar sample at low redshift demonstrates that the $\gamma$ parameter does not show any statistically significant evolution with redshift. This result does not change by using a smaller bin width (at the price of reducing the statistics within each bin) or luminosities instead. 

\begin{figure}[h!]
\centering
\includegraphics[width=\linewidth,clip]{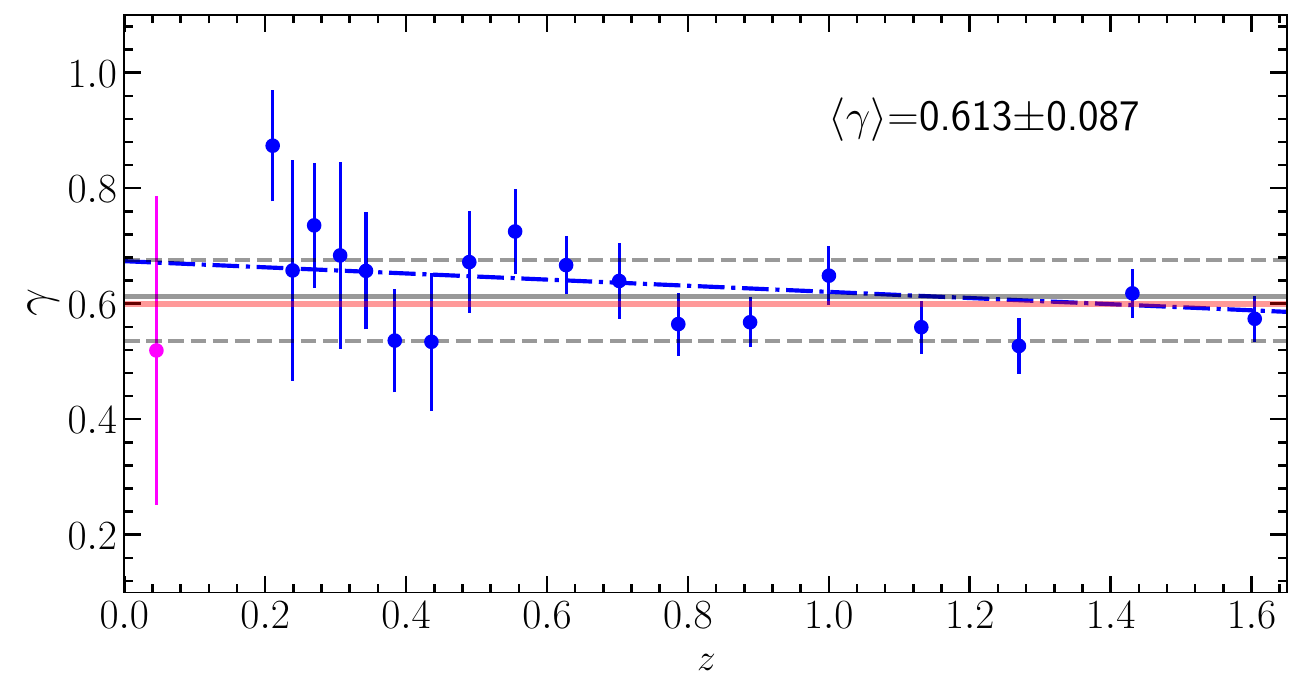}
\caption{Best-fit slope and intrinsic dispersion from the regression analysis of the $\fx-\fo$ relation in narrow redshift intervals from $z$\,$=$\,0.2 to $z$\,$=$\,1.6. A linear fit of $\gamma$ as a function of redshift in the form $\gamma=m\,z+q$ gives $m=-0.052\pm0.018$ and $q=0.673\pm0.026$ (dot-dashed blue line). Host-galaxy contamination at $z$\,$<$\,0.7 mimics a spurious redshift evolution of $\gamma$. For a comparison,
we overplot the resulting fit (magenta point) of the 13 local AGN in the redshift range $0.009$\,$<$\,$z$\,$<$\,0.087, whose flux at 2500 \AA\ is directly evaluated from their UV spectra.}
\label{fig:qsoalllowz}
\end{figure}

We remind that we have a decent number of sources at $z$\,$<$\,0.7 (398 AGN in L20), but these low-redshift quasars are neglected as their UV flux measurements could still be biased by the presence of additional contribution from the host galaxy in the optical. The $\lo$ values for the L20 quasar sample are computed from the photometric SEDs (see their Section 3); in the case of low-luminosity AGN whose contrast with the galaxy is modest, the overall continuum flattens, thus $\lo$ for these galaxies is overestimated. 
The slight shift of the low-redshift (low-luminosity) AGN to higher $\lo$ values for a given X-ray luminosity results in a steepening of the best-fit $\gamma$ in the $\lx-\lo$ plane. This is shown in Figure~\ref{fig:qsoalllowz}, where we performed the same analysis as above but including all the low-redshift AGN in the sample. The slope begins to steepen at $z$\,$<$\,0.7, mimicking a spurious evolution of this correlation parameter with redshift \citep[see, e.g.,][]{li2024}. 

For completeness, we also overplot the resulting fit of the 13 local AGN ($0.009$\,$<$\,$z$\,$<$\,0.087) in the L20 sample. Their $\gamma=0.52\pm0.27$ is consistent within the (large) uncertainties with the expected value of 0.6. The determination of the rest-frame 2500-\AA\ flux in the latter sources relies on the fit of the IUE spectra, which span a wavelength interval (1845--2980 \AA) where the additional contribution of the host galaxy is negligible. Unfortunately, the sample statistics of local AGN with UV spectroscopy covering the rest-frame 2500 \AA\ is not yet sufficient to obtain better constraints.  
Additional UV data of nearby AGN are thus required to improve the determination of the correlation parameters in the redshift range that is key for the cross-calibration of quasars with supernovae.

\begin{figure}[h!]
\centering
\includegraphics[width=\linewidth,clip]{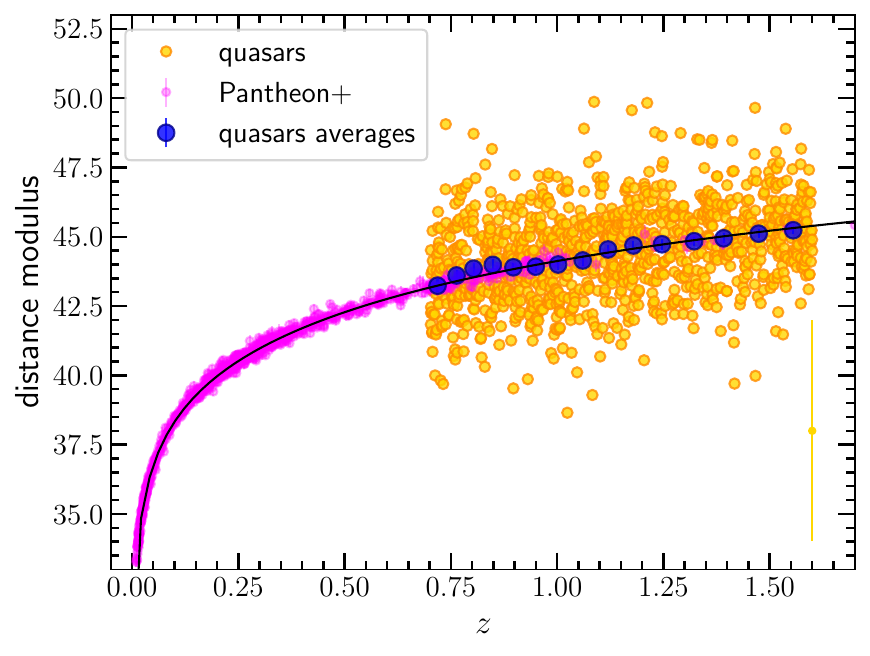}
\caption{Hubble diagram of Pantheon+ supernovae at $z$\,$<$\,1.6 (magenta points) and quasars in the redshift range 0.7\,$<$\,$z$\,$<$\,1.6 (golden points; the average 1$\sigma$ uncertainty on the DM measurements is also shown in the bottom right corner). The blue points are averages of the DM values in narrow redshift bins for quasars only, and are shown for an easier visual comparison. The solid line is the best fit of a flat $\Lambda$CDM model to supernovae and quasars.}
\label{fig:snqsohubblelowz}
\end{figure}

Leveraging on the above results, we can now test the consistency between the distances of supernovae and quasars. 
First, we can fit the quasar sample in the $z$\,$=$\,0.7--1.6 interval with a standard flat $\Lambda$CDM model. To perform the fit, the slope $\gamma$ and the intercept $\beta$ of the relation are left as free parameters, together with $\om$ and an additional parameter $K$ that is included in the definition of the distance modulus, DM:
\begin{equation}
\label{dm}
    {\rm DM} = 5 \log(D_L(z)) + 25 + K,
\end{equation}
with $D_L$ in units of Mpc.
In this way, we can test the cross calibration of quasars with supernovae in the common redshift interval and the consistency of quasar data with the concordance model below $z$\,$=$\,1.6. This test is also very useful to make a direct comparison between the best-fit $\gamma$ parameter and the value obtained from the analysis of the $\fx-\fo$ relation in narrow redshift bins described above.

The results of the simultaneous fit of quasars and supernovae are $\om=0.284 \pm 0.013$, $\gamma=0.605\pm0.015$, $\beta=8.11\pm0.46$, and $K=0.005\pm0.007$. The latter parameter is actually redundant and expected to be relatively small, as the cross-calibration is performed through $\beta$, but it was included to be fully conservative. The parameter $\beta$ is determined by requiring a superposition of the Hubble diagram of quasars to that of supernovae in the common redshift range. The best-fit slope is consistent with the value obtained from the flux--flux analysis of the X-ray to UV relation discussed above, further demonstrating that our assumption of a constant $\gamma$ is robust.
Figure~\ref{fig:snqsohubblelowz} presents the Hubble diagram of quasars and supernovae together with the resulting best cosmological fit to the data. This figure also shows the averages of the DM values in narrow redshift bins for quasars only.  
We emphasise that such averages are not used to perform any cosmological fit, and are presented only for ease of visualization in the Hubble diagram plots. The comparison of the Hubble diagram of quasars with that of supernovae shows that the consistency between these two data sets is remarkable, i.e., the shape of the quasar Hubble diagram is in very good agreement with that of supernovae.

A similar consistency test can be carried out by considering a cosmologically-independent, cosmographic model. The fit of the Hubble diagram was again performed simultaneously for the quasar and supernova data sets in the redshift interval of overlap. We adopted a cosmographic third-order log-polynomial model as in \citet{bargiacchi2021}, leaving the $\gamma$ and $\beta$ parameters free to vary as usual. A third-order polynomial is a conservative choice, as any one of second or higher order would converge rather fast given the limited redshift range. Since the statistical weight of supernovae is largely dominant at $z$\,$<$\,1.6, the shape of the cosmographic model (i.e., the best-fit coefficients of the third-order log-polynomial) will virtually adapt to that data set only with a negligible contribution of quasars, whilst the parameters $\beta$ and $\gamma$ will be predominantly set by the quasar measurements instead. Specifically, the intercept $\beta$ will provide the global calibration between the two data sets and the slope $\gamma$ will adapt its best-fit value to obtain the best possible correspondence between quasars and supernovae. 
The result of the fit is $\gamma=0.601\pm0.015$ and $\beta=8.23\pm0.46$. Again, the slope is in excellent statistical agreement with the cosmology-independent analysis in small redshift bins discussed above. We have also applied a sigma-clipping technique to the quasar data to test whether outliers (at $>$\,3$\sigma$, thus excluding 12 quasars) could introduce any difference in the measurements. The results on $\gamma$ and $\beta$ are fully consistent within the uncertainties (i.e., $\gamma=0.599\pm0.014$ and $\beta=8.32\pm0.45$). 
This is a further proof of the excellent coincidence between the two data sets in the common redshift range.

\section{Testing the evolution of the correlation parameters at high redshift}
\label{sect:gammaevolution}
We now focus on the cosmological analysis of the Hubble diagram for quasars only, aiming at addressing the case of an evolution of the normalization $\beta$ and of the slope $\gamma$ in the redshift domain inaccessible to supernovae. We will do so in three steps. We first show that in our method a complete degeneracy is present between a redshift evolution $\beta(z)$ and the shape of the distance--redshift relation. As a consequence, contrary to the case of $\gamma$, the issue of a redshift evolution of $\beta$ can be only evaluated in the light of our current understanding of the X-ray to UV relation itself. We then provide several observational results supporting the hypothesis of no redshift evolution of the correlation parameters. Finally, we repeat the analysis of the $\fx-\fo$ relation extending to $z$\,$>$\,1.6 and adopting diffetent sizes of the redshift bins.

\subsection{Degeneracy between the distance--redshift and $\beta$--redshift relations}
\label{beta-z}
We start by arguing that an intrinsic degeneracy exists between the distance--redshift relation and a possible evolution with redshift of the intercept $\beta$, and that this degeneracy is present irrespective of the cosmological model.  
Based on equation (\ref{betahat}), it is straightforward to demonstrate that, if we assume a `true' cosmological model and a non-evolving X-ray to UV relation as in equation (\ref{lxlo}), such a relation can be reproduced within any alternative cosmological model, provided that the parameter $\beta$ evolves as:
\begin{equation}
\label{betaz}	
\beta(z)=\beta-(2\gamma-2)\log\frac{D_L^*(z)}{D_L(z)},
\end{equation}
where $D_L(z)$ and $D_L^*(z)$ are the luminosity distances computed with the `true' model and with the alternative one, respectively. This formal equivalence is an intrinsic feature of the standard-candle method. Since it holds precisely at all redshifts, it implies that if we assume an analytic function $\beta(z)$ with enough flexibility to reproduce the redshift-dependent term in equation~(\ref{dm}), it is impossible to disentangle the correct cosmological model from a fitting procedure of observational data, even for perfect standard candles: all the fits with any possible cosmological model will always produce the same likelihood value. Besides the formal correctness of the above statement, it is easy to verify that even considering cosmological models very different from each other, the redshift-dependent term in equation~(\ref{dm}) has typically a simple shape that can be readily reproduced by an ordinary analytical function containing only a couple of free parameters. 

\begin{figure}[t!]
\centering
\includegraphics[width=\linewidth,clip]{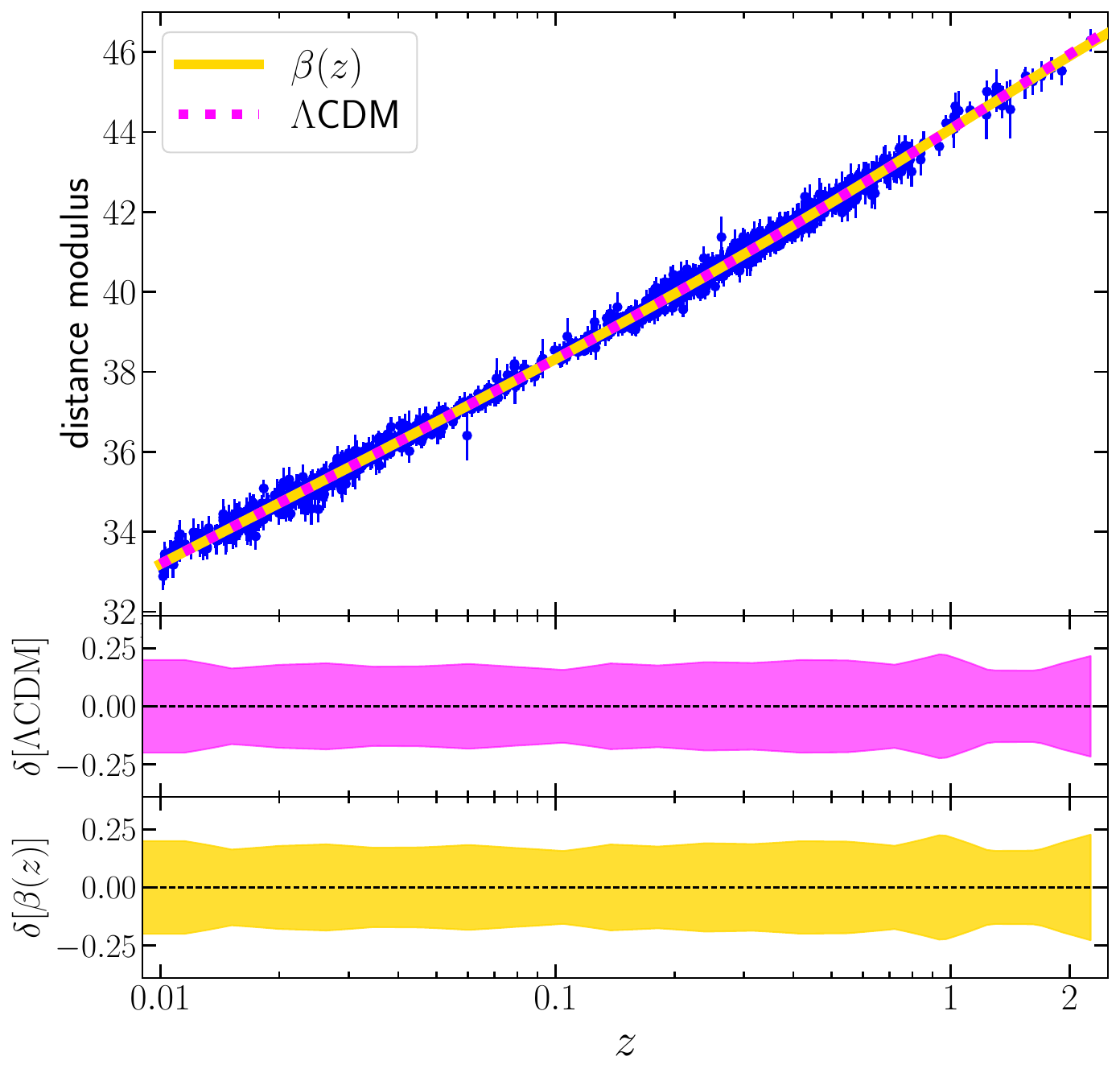}
\caption{Top panel: Hubble diagram of the Pantheon+ supernovae sample, fitted with an open $\Lambda$CDM model with $\om$ and $\Omega_\Lambda$ as free parameters, and with a $\Lambda$CDM model with fixed $\om$\,=\,1 and $\Omega_\Lambda$\,=\,0 and allowing for a redshift evolution of the absolute zero-point magnitude of supernovae. Bottom panels: dispersion with respect to the best fit for the two models.}
\label{fig:joke}
\end{figure}

To further elucidate this point, we show the effects of the degeneracy using the Pantheon+ supernova sample, which is widely accepted as a reliable distance indicator. We remind that within our method, the quantity that contains the cosmologically relevant information is the parameter $\hat{\beta}$ in equation~(\ref{betahat}), i.e., the absolute calibration of the relation in {\em flux} units at a given redshift.  
In this respect, assuming a redshift evolution of $\beta$ in our analysis is {\em formally} equivalent to assuming a redshift dependence of the zero-point reference magnitude of type Ia supernovae; discussing whether these assumptions are justified is beyond the scope of this work, but we want to elaborate on their formal consequences. 
We first fitted the Pantheon+ sample in the standard way, i.e., assuming the distance moduli as estimated by \citet{Scolnic2022} and adopting a flat $\Lambda$CDM model. We obtained the well-known results presented in \citet{Brout2022}, and we computed the dispersion from the best fit as a function of redshift. The results are shown in Figure~\ref{fig:joke}. We then repeated the fit assuming a fixed $\Lambda$CDM model with $\om$\,=\,1 and $\ol$\,=\,0, and an evolution of the supernova zero-point magnitude of the form $a_1\exp(a_2 z)$, with $a_1$ and $a_2$ as free parameters. Despite the rather extreme values chosen for the cosmological model, we were able to obtain a fit perfectly equivalent to the first one, which is also shown in Figure~\ref{fig:joke}.
The best-fit values of the free parameters are fully reasonable ($a_1$\,$\sim$\,$-0.3$; $a_2$\,$\sim$\,$-0.5$). 

We conclude that there is no way to rule out the latter scenario based on the cosmological analysis, although we had purposely chosen an unreasonable cosmological model. It is obviously possible to dismiss the matter-only model based on other cosmological probes, yet we can safely discard this solution in the context of an analysis solely based on supernovae simply because we believe we have enough observational evidence and physical understanding about supernovae Ia to rule out the luminosity evolution implied by the best-fit values of the parameters $a_1$ and $a_2$. This exercise shows that any departure of the assumed cosmological model from the correct (and unknown) one can be compensated for by a suitable, albeit artificial evolution of some physical property of the standard candle at issue. For this reason, one cannot rely on a cosmological analysis to determine the ultimate standardisability of a given data set.
The analogy with quasars is straightforward: it is not possible to reject, nor to confirm any evolution of the absolute calibration of the X-ray to UV relation based on a cosmological analysis. Therefore, the reliability of the method entirely depends on the observational evidence supporting the non-evolution hypothesis.

\subsection{Observational evidence for a non-evolving X-ray to UV relation}
\label{Observational evidence for a non-evolving X-ray to UV relation}

While the analysis of the $F_{\rm X}-F_{\rm UV}$ relation in narrow redshift intervals confirms the redshift-independence of $\gamma$, we do not have any such direct test for $\beta$. 
However, a first, fundamental argument against the possible evolution of $\beta$ with redshift is the {\it universal} nature of black-hole accretion. It is well established that the accretion process is one and the same irrespective of black-hole mass, and that AGN are a scaled-up version of stellar-mass black holes \citep[e.g.,][]{mchardy2006}. Likewise, once a black hole starts to steadily accrete matter, and no other physical mechanisms are involved (e.g., strong outflows, or jets), there is no reason to believe that accretion physics should be any different at low and high redshifts. Indeed, no theoretical model has yet proposed, or even postulated, a variation in the accretion process only based on redshift. In this wake, there is ample indirect observational evidence that $\beta$ must be independent of redshift:

\begin{enumerate}
    \item We observe an excellent agreement between the Hubble diagrams of quasars and supernovae in the common redshift range. Despite the large scatter of quasars, there is no systematic offset, nor any trend that may suggest a possible evolution of $\beta$ at least up to $z$\,$\simeq$\,1.5. An evolving $\beta$ at higher redshifts thus appears like a sort of fine-tuning required to reconcile the discrepancy observed in the Hubble diagram of high-redshift quasars with the $\Lambda$CDM predictions.
    \item $\Delta \alpha_{\rm OX}$ (i.e., the difference between the observed and the predicted $\alpha_{\rm OX}$) is independent on redshift. For any given luminosity, the parameter $\alpha_{\rm OX}$, defined as $0.384\log(\lx/\lo)$ after \citet{avnitananbaum79}, has been found to be constant across the entire redshift range probed so far (see, for instance, Figure 4 in \citealt{vito2019}).
    \item The similarity between the average optical/UV spectrum of standard blue quasars at $z$\,$\lesssim$\,1.6 \citep[e.g.,][]{vandenberk2001} and that observed in the highest-redshift quasars ($z$\,$\gtrsim$\,7; see, e.g., \citealt{Mortlock2011,banados2018}) is striking. Such a spectral homogeneity implies that the underlying physical process is the same across a wide redshift range. This is quantitatively confirmed by the analysis presented in \citet{trefoloni2024}, where we reveal no evolution of the quasar properties (e.g., the nuclear continuum slope) when the sources are selected to be blue, with little radio emission and no broad absorption lines, ensuring the absence of strong jets and/or outflows.
    \item We individually analysed the X-ray and UV spectra of the 130 quasars at $z$\,$>$\,2.5 in L20 in \citet{sacchi2022}, finding that there is no difference in either band between the nuclear continuum and overall spectral properties of high-redshift quasars and those of their low-redshift analogues.
    \item The correlation between $\lx$ and $\lo$ remains tight, if not slightly tighter, even when different proxies for the X-ray and UV emission are adopted 
    \citep{signorini2023}. When the residual contributions to the observed dispersion are investigated (i.e., those that cannot be removed through a careful sample selection, like variability, or accretion disc inclination), the intrinsic dispersion in the X-ray to UV relation turns out to be $<$\,0.06 dex \citep{signorini2024}.
\end{enumerate}

All these pieces of observational evidence support the non-evolution of the $\gamma$ and $\beta$ parameters with redshift in typical quasars, in keeping with the notion that accretion physics is the same at low and high redshifts. 
Even so, we always need to check whether the $F_{\rm X}-F_{\rm UV}$ relation for the adopted quasar sample shows any trend with redshift before performing any cosmological analysis. 
This issue was addressed in depth in previous works from our group (see Section 8 in L20); yet, below we further demonstrate that our method to test for the redshift evolution (or lack thereof) of  the correlation parameters is robust.

\begin{figure}[t!]
\centering
\includegraphics[width=\linewidth,clip]{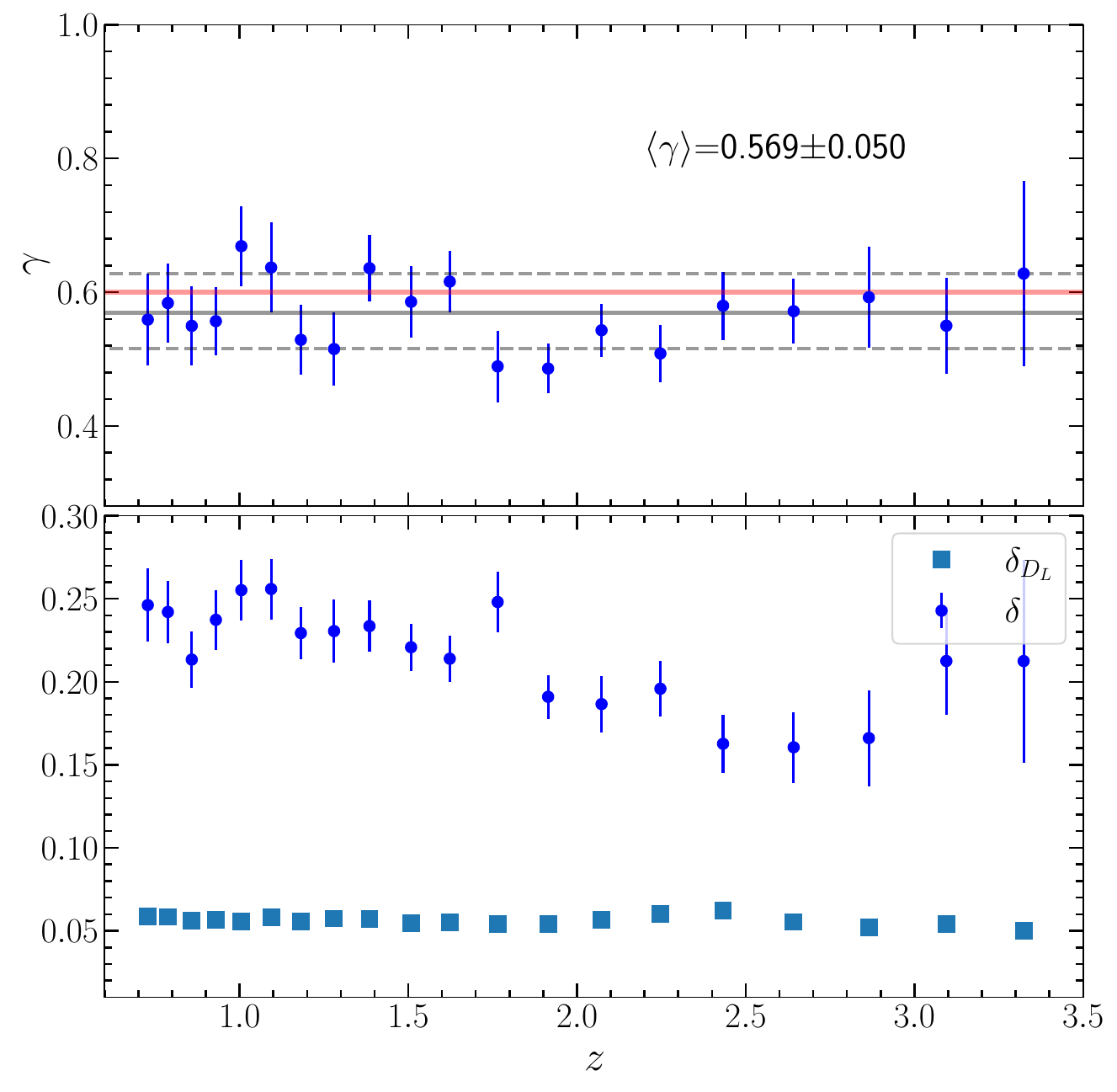}
\caption{Redshift evolution of the slope of the $F_{\rm X}-F_{\rm UV}$ relation in narrow redshift bins (top panel). To perform the regression fit, X-ray and UV fluxes are normalized to $10^{-31}$ and $10^{-27}$ erg s$^{-1}$ cm$^{-2}$ Hz$^{-1}$, respectively. The data points are plotted at the average redshift within the interval. Error bars represent the 1$\sigma$ uncertainty on the mean in each bin. The grey solid and dashed lines are the mean and 1$\sigma$ uncertainty range on the slope ($\gamma$). The red line marks $\gamma$\,$=$\,0.6. The bottom panel shows the dispersion ($\delta$, blue filled circles) along the best fit of the $F_{\rm X}-F_{\rm UV}$ relation and the dispersion of the distance distribution ($\delta_{DL}$\,$\simeq$\,0.06, square symbols) in each bin. See Section~\ref{subsect:gamma-z} for details.}
\label{fig:gamma-dz}
\end{figure}
\begin{figure}[h!]
\centering
\includegraphics[width=\linewidth,clip]{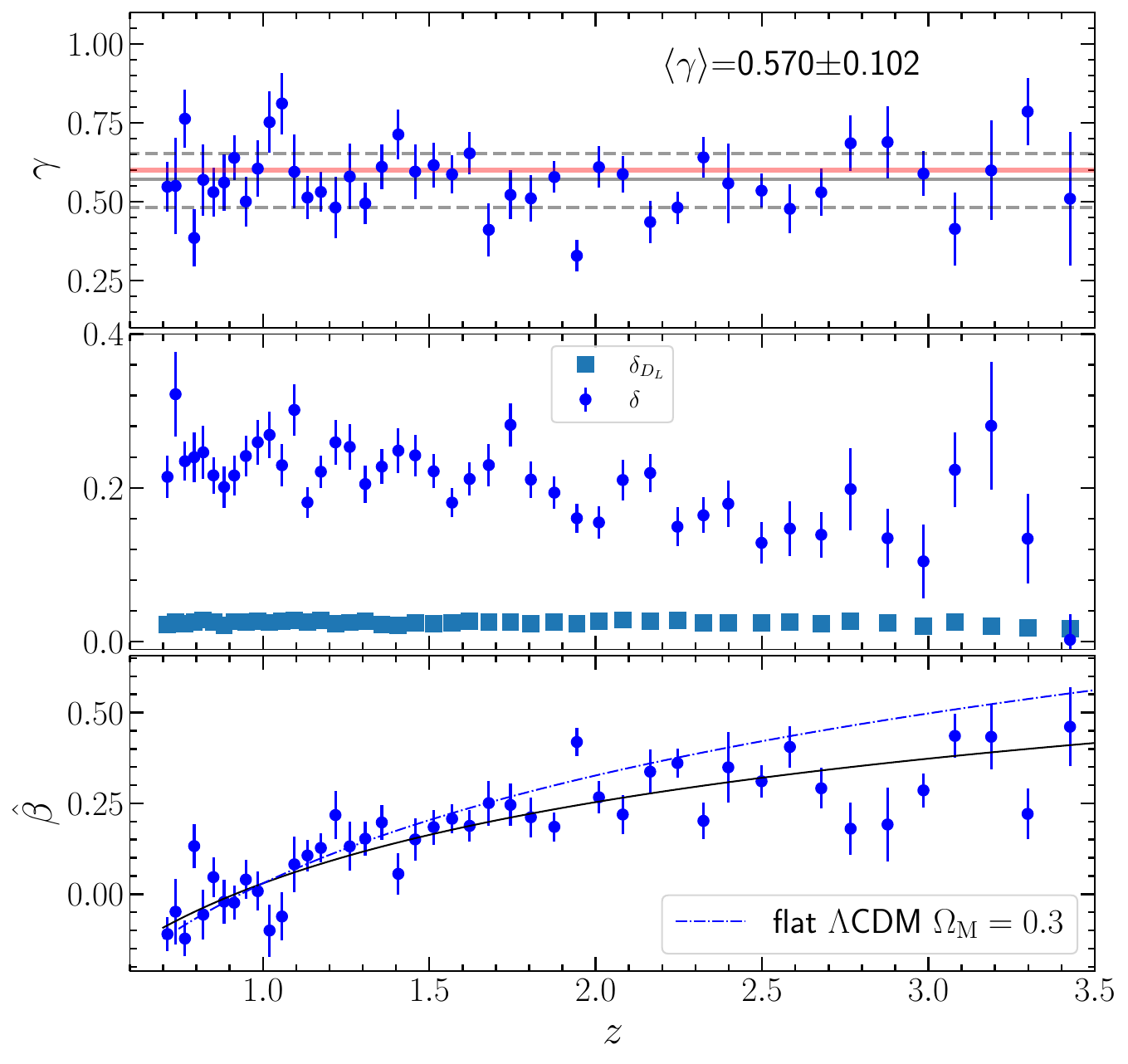}
\caption{Same as for Figure~\ref{fig:gamma-dz} with a much finer binning within the redshift range 0.7$-$3.5. The bottom panel presents $\hat\beta$ from equation \ref{betahat}} as a function of redshift. The solid black line is a second-order cosmographic fit of the data, for which a Gaussian prior is considered on both $\gamma$ and $\beta$ (i.e., $\gamma=0.600\pm0.005$ and $\beta=8.00\pm0.25$). The dot-dashed line is a flat $\Lambda$CDM model with $\Omega_{\rm M}$ fixed to 0.3, $\gamma$ and $\beta$ fixed to 0.6 and 8.0, and normalised to the cosmographic fit at $z$\,$=$\,1.
\label{fig:testbetahat}
\end{figure}

\subsection{Non-evolution of $\gamma$ at $z$\,$>$\,1.6}
\label{subsect:gamma-z}
\noindent
Following from Section~\ref{sect:qso-sn comparison}, we divided the sample of quasars in narrow redshift bins within the interval 0.7--3.5, with a logarithmic step whose width is set by two criteria: each redshift interval must be characterised by enough statistics (the number of objects is $>$\,15 in each bin), and by a dispersion in distances smaller than the dispersion in the $\fx-\fo$ relation. 
As we have already mentioned, this test serves not only to verify that $\gamma$ does not evolve with redshift, but also to ascertain its {\it intrinsic} value, independent from any cosmological model.
Figure~\ref{fig:gamma-dz} shows the slope resulting from the regression analysis of the $F_{\rm X}-F_{\rm UV}$ relation computed in narrow redshift bins as a function of redshift. The bottom panel shows the values for the dispersion ($\delta$) along the best fit of the $F_{\rm X}-F_{\rm UV}$ relation compared to the dispersion of the luminosity distances, i.e., $\delta_{D_L}=\sigma(D_L^2)/\langle D_L^2\rangle$, where $\sigma(D_L^2)$ and $\langle D_L^2\rangle$ are, respectively, the standard deviation and the mean of the distribution of the squared distances within each bin, which have been considered to be more conservative.   
The dispersion associated with the distance distribution is indeed much smaller than the overall dispersion of the $F_{\rm X}-F_{\rm UV}$ relation in each bin.\footnote{The dispersion of the $\fx-\fo$ relation is always higher also compared to the relative difference between the maximum and minimum distances (squared) within the bin, i.e., $(D_{L, \rm{max}}^2-D_{L, \rm{min}}^2)/\langle D_L^2\rangle$, independently on the exact cosmology.}

Figure~\ref{fig:testbetahat} presents the redshift evolution of the slope of the $F_{\rm X}-F_{\rm UV}$ relation adopting a finer binning. As expected, the average slope is fully consistent with the one shown in Figure~\ref{fig:gamma-dz}, although the uncertainties are larger as the number of sources within each bin is much smaller, especially at $z$\,$>$\,2.5. Therefore, the average $\gamma$ value is not dependent upon the binning choice, as long as the distance dispersion within the bin is such that $\delta_{D_L}$\,$<$\,0.1, that is, lower than the minimum observed dispersion reasonably expected for the X-ray to UV relation (cf. \citealt{sacchi2022}). Here, we also show the resulting $\hat\beta$ as a function of redshift, which displays the expected evolution of the luminosity distance with redshift expressed in equation \ref{betahat}. We fit the data points with a second-order cosmographic fit of the form:
\begin{equation}
    D_L= \frac{c\ln(10)}{H_0}\, (x + a_1 x^2), 
\end{equation}
where $x$\,=\,$\log(1+z)$, $a_1$ is the polynomial coefficient, $c$ is the speed of light, and $H_0$ is fixed to 70 km s$^{-1}$ Mpc$^{-1}$. The second-order cosmographic fit is presented for visualization purposes only, as $\hat\beta$ is not calibrated, hence the best-fit values of the model have no cosmological meaning. A Gaussian prior is considered for both $\gamma$ and $\beta$ in the cosmographic fit (i.e., $\gamma=0.600\pm0.005$ and $\beta=8.00\pm0.25$). We also plot a flat $\Lambda$CDM model with $\Omega_{\rm M}$ fixed to 0.3, and $\gamma$ and $\beta$ fixed to 0.6 and 8.0, respectively. This model is normalised to the cosmographic fit at $z$\,=\,1. 
This (qualitative) comparison shows that, whilst there is consistency of the distances below $z$\,$\approx$\,1.7, the data points deviate from the predictions of a flat $\Lambda$CDM model at higher redshifts. 

Again, such a discrepancy can be interpreted in two alternative ways: since we have demonstrated that $\gamma$ does not show any systematic trend, either there is an evolution of $\beta$ with redshift, or a flat $\Lambda$CDM model is not a good representation of the data.
However, the Hubble diagram of quasars is in excellent agreement with that of supernovae in the common redshift range (see Figure~\ref{fig:snqsohubblelowz}), hence also $\beta$ is constant at least up to $z$\,$\sim$\,1.5. Moreover, all the physical and spectral properties of the quasars selected at $z$\,$>$\,2.5 do not differ from those of lower redshift AGN \citep[i.e.,][]{sacchi2022,trefoloni2024}. We therefore conclude that this deviation is most likely a limitation of the flat $\Lambda$CDM model.

\begin{figure*}[h!]
\centering
\includegraphics[width=0.49\linewidth,clip]{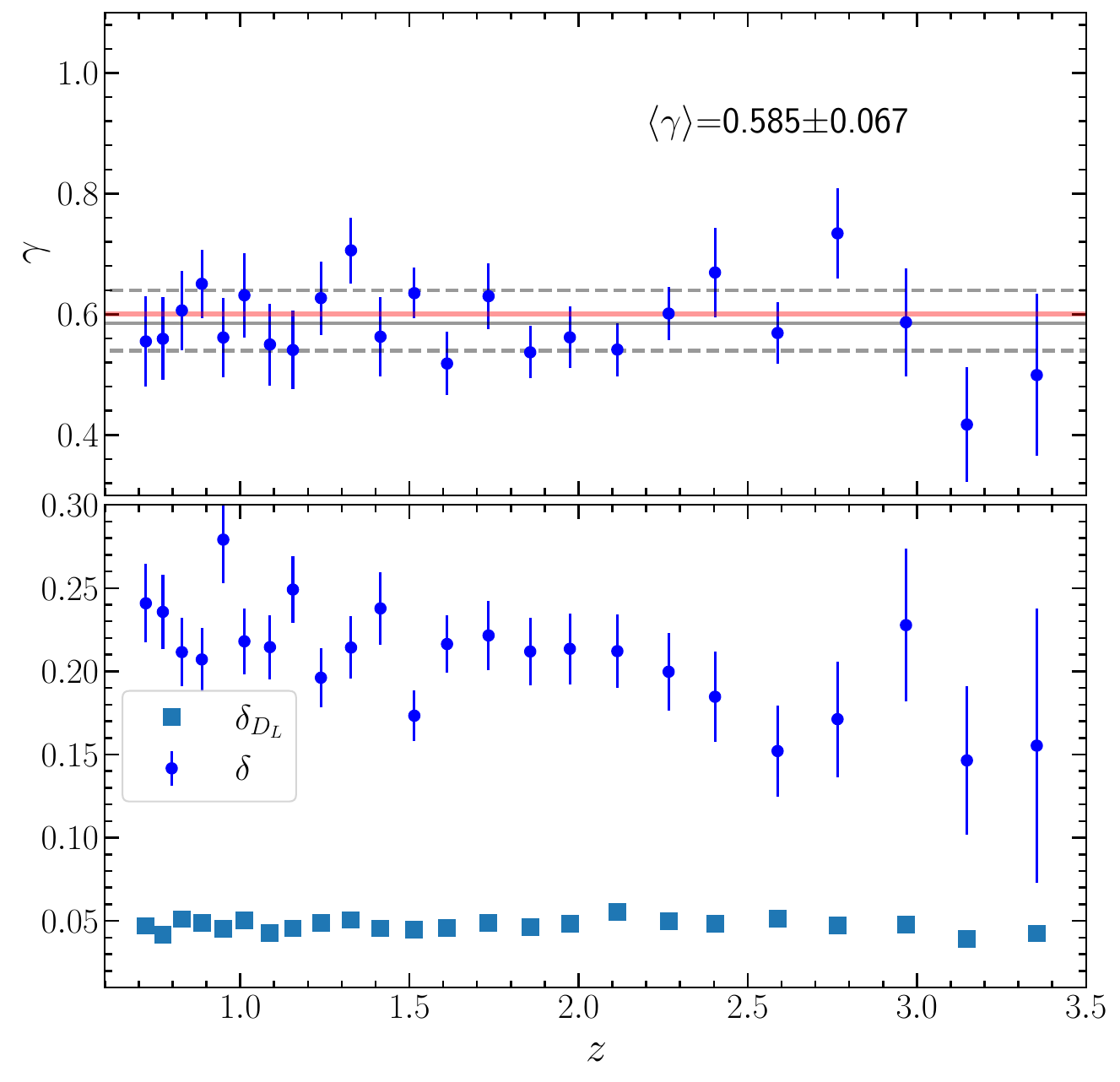}
\includegraphics[width=0.49\linewidth,clip]{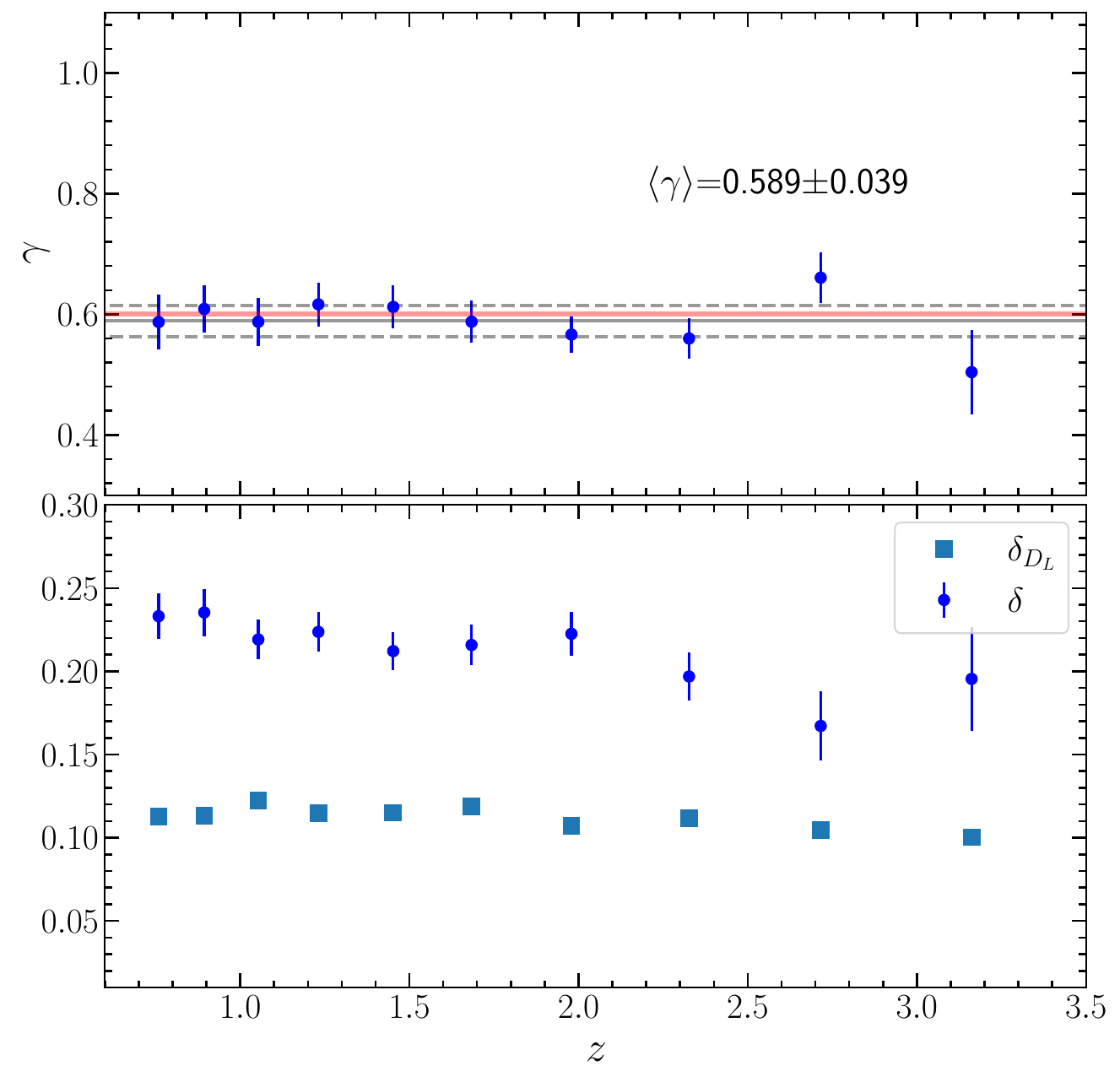}
\caption{Same as Figure~\ref{fig:gamma-dz} for the simulated quasar sample 1 (2023 sources) with two different binning criteria: the one on the left is the same as for the real data (24 intervals, $\delta_{DL}$\,$\simeq$\,0.06), while the one on the right considers a coarser binning (10 intervals, $\delta_{DL}$\,$\simeq$\,0.1). In both cases, we can retrieve the input value of $\gamma$ within uncertainties, supporting the assumption that $\gamma$ is constant in the analysis of real data independently from any cosmological model (here $\om$\,=\,0.1 and $\ol$\,=\,0.9 were assumed)}.
\label{fig:simcheck-gammabin}
\end{figure*}

\section{On the degeneracy between evolutionary and cosmological models}
\label{sect:Simulations}
We now expand on the analysis presented in Section \ref{beta-z} and run a series of simulations to illustrate that it is not possible to draw any conclusions regarding the reliability of our technique purely based on the comparison of cosmological models \citep[e.g.,][]{KR2020,KR2021,KR2022}, even when accounting for the possible evolution of the correlation parameters $\gamma$ and $\beta$ with redshift. For such an approach to work, one would have to know the `true' cosmological model, whereas even in the case of the widely accepted $\Lambda$CDM parameterisation, it has been demonstrated by several groups that this model has pitfalls. The logical fallacy lies in the fact that a cosmological model (or a suite of very different ones) cannot be employed to validate the {\it standardisability} of a specific cosmological probe, or data set. 
The determination of the correlation parameters $\gamma$ and $\beta$ can thus be obtained only from observational evidence, independently of any cosmological model, as we have discussed in Section~\ref{sect:gammaevolution}.

\subsection{The mock quasar sample}
We simulated a quasar sample starting from the redshift and UV flux distributions of the L20 data set, where only $z$\,$>$\,0.7 quasars are considered (i.e., 2023 quasars within the redshift interval 0.703\,$<$\,$z$\,$<$\,7.540, with an average redshift of $\simeq$\,1.6). With respect to the observed distributions, we introduced a small random offset to the value of both redshifts and UV fluxes (in log units), assuming a Gaussian distribution centred at zero with dispersion $\sigma_z$\,=\,0.001
and $\sigma_{F_{\rm UV}}$\,=\,0.01.  
From the simulated redshift distribution we computed the luminosity distances adopting $\om$\,=\,0.1 and $\ol$\,=\,0.9, whose values were chosen to be very different from the ones of the $\Lambda$CDM model or its simplest extensions. By combining the resulting distances and the simulated UV fluxes, we obtained mock UV luminosities, from which we subsequently derived the corresponding X-ray luminosities by assuming $\gamma$\,=\,0.6, $\beta$\,=\,8.0, and a dispersion $\delta$ of the $\lx-\lo$ relation of 0.24 dex (see, e.g., \citealt{lr16}). We then converted X-ray luminosities back into fluxes, still in the framework of the above-mentioned cosmology. Uncertainties were assigned to both X-ray and UV fluxes by applying, again, to the observed ones a small random offset drawn from a Gaussian distribution. 
In order to evaluate the effects of sample statistics on the results, we considered two samples: one with exactly the same number of quasars as in the real data set (2023, sample 1), and another where we increased the statistics by a factor of 10 (i.e., 20230 sources, sample 2). 

Before embarking upon the cosmological analysis, we have also examined whether the determination of the average correlation parameter $\gamma$ in this clearly exotic cosmology presents any anomaly or bias. For extra caution, we adopted two different widths of the redshift bins.
Figure~\ref{fig:simcheck-gammabin} presents the evolution of the slope of the $F_{\rm X}-F_{\rm UV}$ relation with redshift for the simulated quasar sample 1 (2023 sources). The plot in the left panel is obtained with the same number of intervals as in the real data (i.e., 24 intervals, $\delta_{DL}$\,$\simeq$\,0.06), while the plot in the right panel displays a coarser binning (i.e., 10 intervals, $\delta_{DL}$\,$\simeq$\,0.1). In both cases, we can retrieve the input $\gamma$ value within uncertainties, thus confirming that we can safely assume the same $\gamma$ for both the $\lx-\lo$ and the $\fx-\fo$ relations (i.e., $\gamma_L$\,=\,$\gamma_F$ in the formalism of \citealt{petrosian2022}) when applying this technique to the analysis of real data.
\begin{table*}[h!]
\caption{\label{tbl:nonflatlcdm} Marginalized one-dimensional best-fitting parameters with 1$\sigma$ confidence intervals for the simulated quasar sample 1 (2023 sources).}
\centering
\begin{tabular}{lccccrrrr}
\hline\hline
Model & $\om$ & $\ol$ & $\gamma$ & $\beta$ & $\gamma_0$ & $\gamma_1$ & $\beta_0$& $\beta_1$\\
\hline
1 & $0.13\pm0.07$ & $0.43\pm0.30$ & $0.584\pm0.011$ & $8.44\pm0.34$ & - & - & - & - \\
2 & 0.3 & 0.7 & $0.562\pm0.009$ & $9.08\pm0.27$ & - & - & - & - \\
3 & $0.42\pm0.28$ & $0.68\pm0.42$ & - & - & $0.560\pm0.032$ & $0.010\pm0.011$ & $9.23\pm0.98$ & $-0.33\pm0.36$ \\
4 & 0.3 & 0.7 & - & - & $0.568\pm0.026$ & $0.007\pm0.009$ & $8.98\pm0.80$ & $-0.23\pm0.30$ \\
5 & 0.9 & 0.1 & $0.545\pm0.009$ & $9.49\pm0.27$ & - & - & - & - \\
6 & 0.9 & 0.1 & - & - & $0.541\pm0.027$ & $0.015\pm0.010$ & $9.76\pm0.83$ & $-0.51\pm0.30$ \\
\hline
\end{tabular}
\tablefoot{The simulation assumes $\gamma$\,=\,0.6 and $\beta$\,=\,8.0, with a dispersion of the $\lx-\lo$ relation of $\delta$\,=\,0.24. The distances have been computed assuming a spacially flat $\Lambda$CDM model with $\om$\,=\,0.1 and $\ol$\,=\,0.9 (see Section~\ref{sect:Simulations} for details). The $\Lambda$CDM model has $\om$ and $\ol$ free to vary for models 1 and 3, whilst models 2 and 4 have the matter and energy parameters fixed to the values 0.3 and 0.7, respectively. Models 3 and 4 assume an evolution with redshift of both slope and intercept of the X-ray to UV relation in the form: $\gamma(z)=\gamma_0+\gamma_1(1+z)$ and $\beta(z)=\beta_0+\beta_1(1+z)$.
In all models, $\Omega_{k}$ is fixed to zero (spacially flat case) and $\Omega_{r}=2.47\times 10^{-5}\,h^{-2}$, with $h=0.7$ km s$^{-1}$ Mpc$^{-1}$, $\Omega_{r}=1-\om-\ol$, and $\ol<0.99+1.8\,\om$. The latter condition has been enforced to exclude solutions with no Big Bang. Models 5 and 6 are the same as 2 and 4, respectively, but with matter and energy parameter values fixed to $\om=0.9$ and $\ol=0.1$. Model 3 is a sanity check to test what happens when the cosmological parameters are free to vary and an evolution of the correlation parameters is allowed.}
\end{table*}

\begin{table*}[h!]
\caption{\label{tbl:nonflatlcdm-big}Same as Table~\ref{tbl:nonflatlcdm} for the simulated quasar sample 2 (20230 quasars; see Section~\ref{sect:Simulations} for details).}
\centering
\begin{tabular}{lccccrrrr}
\hline\hline
Model & $\om$ & $\ol$ & $\gamma$ & $\beta$ & $\gamma_0$ & $\gamma_1$ & $\beta_0$& $\beta_1$\\
\hline
1 & $0.10\pm0.02$ & $0.93\pm0.05$ & $0.596\pm0.004$ & $8.12\pm0.12$ & - & - & - & - \\
2 & 0.3 & 0.7 & $0.576\pm0.003$ & $8.65\pm0.09$ & - & - & - & - \\
3 & $0.10\pm0.04$ & $1.07\pm0.08$ & - & - & $0.625\pm0.013$ & $-0.011\pm0.005$ & $7.26\pm0.40$ & $0.32\pm0.14$ \\
4 & 0.3 & 0.7 & - & - & $0.579\pm0.008$ & $0.005\pm0.003$ & $8.61\pm0.26$ & $-0.17\pm0.09$ \\
5 & 0.9 & 0.1 & $0.559\pm0.003$ & $9.07\pm0.09$ & - & - & - & - \\
6 & 0.9 & 0.1 & - & - & $0.555\pm0.009$ & $0.130\pm0.003$ & $9.32\pm0.28$ & $-0.43\pm0.11$ \\
\hline
\end{tabular}
\end{table*}

\subsection{Results for spacially flat $\Lambda$CDM models}
\label{subsect:spaciallyflatlCDMmodel}
\begin{figure*}[h!]
\centering
\includegraphics[width=\linewidth,clip]{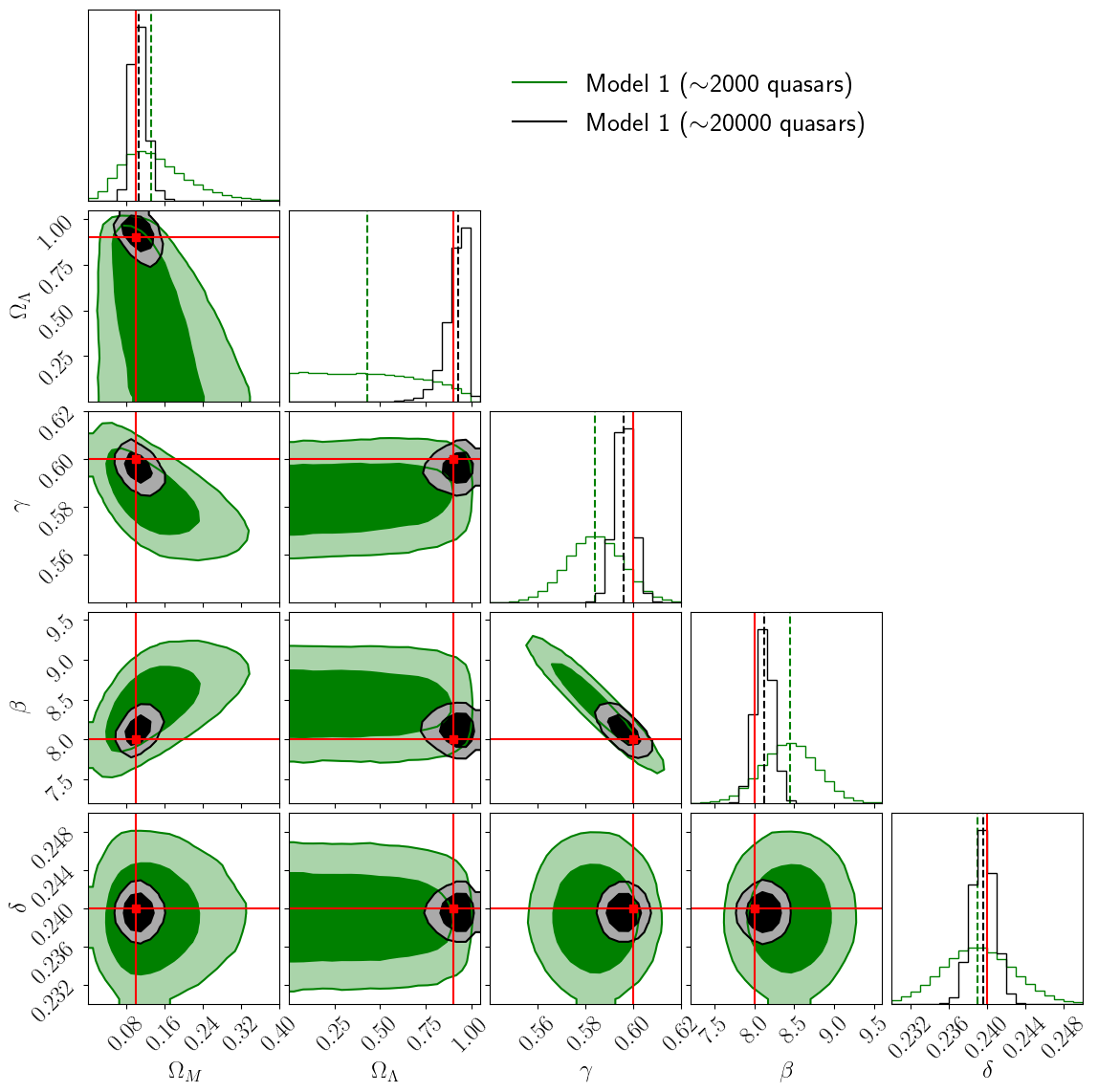}
\caption{Results of the cosmological fit for both sample 1 (green) and sample 2 (black) for a spacially flat $\Lambda$CDM with $\om$ and $\ol$ free to vary. The correlation parameters $\gamma$ and $\beta$ (as well as the dispersion $\delta$) of the X-ray to UV relation are also free to vary (see Section~\ref{subsect:spaciallyflatlCDMmodel} for details). Contours are at 68\% and 95\% confidence levels. The input values of both the cosmological and correlation parameters are marked by the red square. All the input parameters considered to build the mock quasar samples are retrieved within the uncertainties.}
\label{fig:simcheck}
\end{figure*}

We first fit both sample 1 and sample 2 with a $\Lambda$CDM model where $\om$ and $\ol$ are free to vary (model 1). The curvature $\Omega_{k}$ is set to zero (i.e., spacially flat case) and the radiation density is fixed to the observed value $\Omega_{r}=2.47\times 10^{-5}\,h^{-2}$, with a Hubble parameter $h=0.7$ km s$^{-1}$ Mpc$^{-1}$. We also imposed the following two conditions: $\Omega_{r}=1-\om-\ol$ and $\ol<0.99+1.8\,\om$, the latter aimed at excluding solutions with no Big Bang. The results of the fit are shown in Figure~\ref{fig:simcheck}.
The marginalized one-dimensional best-fitting parameters with $1\sigma$ confidence intervals are summarised in Tables~\ref{tbl:nonflatlcdm} and \ref{tbl:nonflatlcdm-big} for samples 1 and 2, respectively. 
As expected, we can retrieve the input values of both the cosmological model and the correlation parameters within the uncertainties, with just a marginal offset of $\ol$ for sample 1. This effect is due to the low sample statistics, especially at $z$\,$<$\,1, where the energy density dominates. When sample 2 is considered, the value of $\ol$ is better constrained thanks to the larger sample statistics in the redshift range 0.7--1.6. 

\begin{figure}[h!]
\centering
\includegraphics[width=\linewidth,clip]{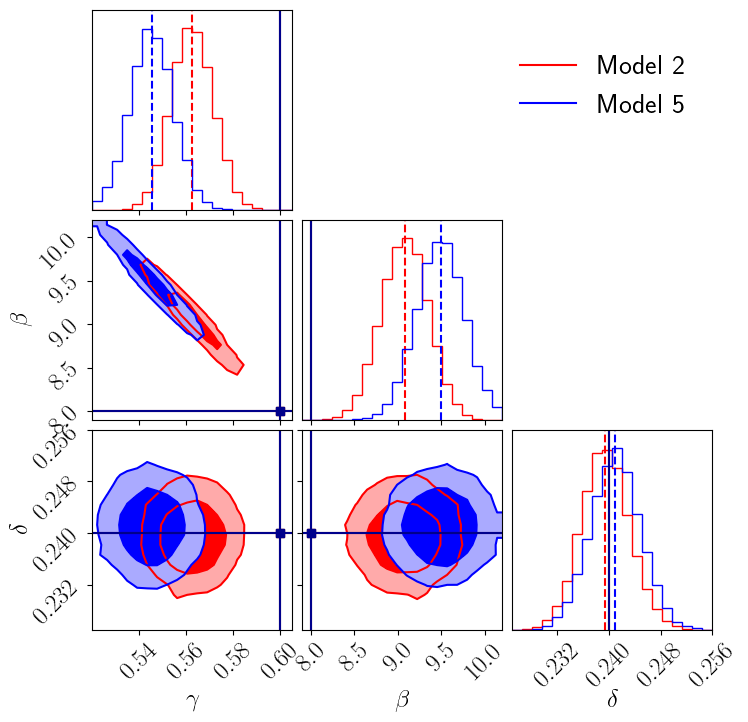}
\caption{Results of the cosmological fit for sample 1 in a spacially flat $\Lambda$CDM model with matter and energy density parameters fixed to the values $\om$\,=\,0.3 and $\ol$\,=\,0.7 (red, model 2 in Table~\ref{tbl:nonflatlcdm}) and $\om=0.9$ and $\ol=0.1$ (blue, model 5 in Table~\ref{tbl:nonflatlcdm}). Contour levels are at 68\% and 95\% confidence. The input values of correlation parameters are marked by the black square. See Section~\ref{subsect:spaciallyflatlCDMmodel} for details.}
\label{fig:simcheck-fix}
\end{figure}
\begin{figure}[h!]
\centering
\includegraphics[width=\linewidth,clip]{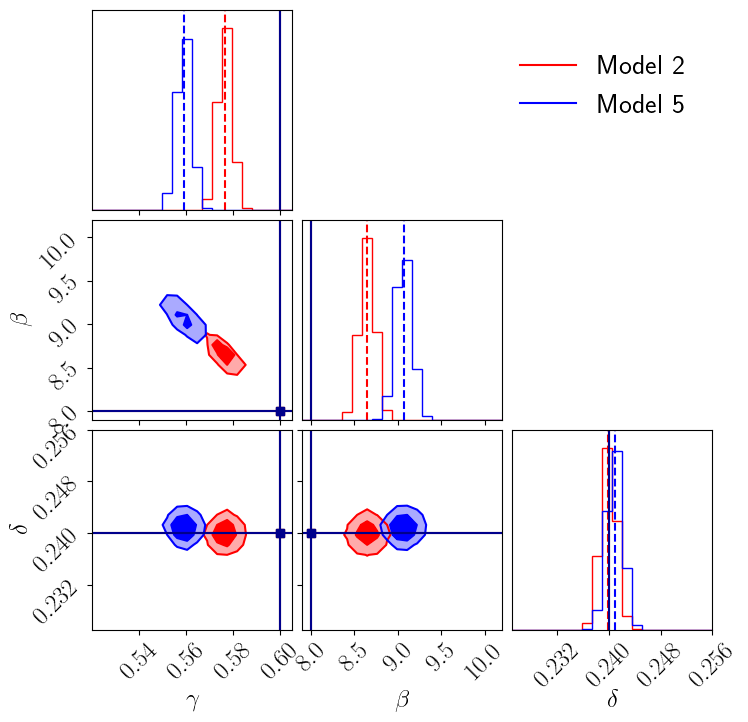}
\caption{Same as in Figure~\ref{fig:simcheck-fix} for  sample 2.}
\label{fig:simcheck-fix-big}
\end{figure}

We then fit both samples with the same cosmological model but with the matter and energy density parameters fixed to the values $\om$\,=\,0.3 and $\ol$\,=\,0.7 (model 2), and $\om$\,=0.9 and $\ol$\,=\,0.1 (model 5). 
The results of the relative cosmological fits are shown in Figures~\ref{fig:simcheck-fix} and \ref{fig:simcheck-fix-big} for samples 1 and 2, respectively. The correlation parameters $\gamma$ and $\beta$ deviate from the input values at more than the $5\sigma$ statistical level, and this effect is even more significant when $\om$ and $\ol$ are fixed to values that substantially differ from the `true' ones, as is the case for $\om$\,=\,0.9 and $\ol$\,=\,0.1. 

\begin{figure*}[h!]
\centering
\includegraphics[width=\linewidth,clip]{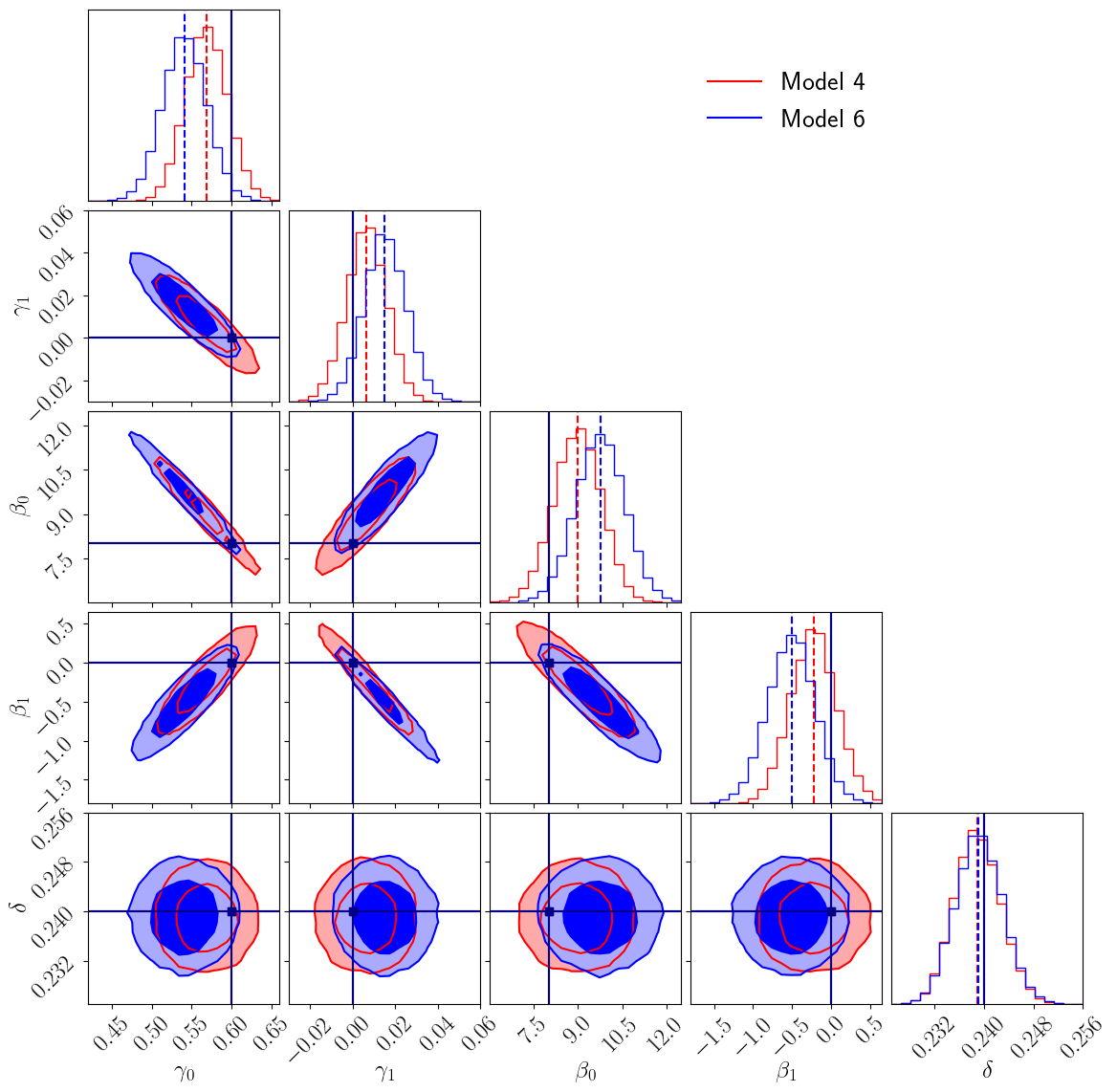}
\caption{Results of the cosmological fit for sample 1 in a spacially flat $\Lambda$CDM with matter and energy density parameters fixed to the values $\om$\,=\,0.3 and $\ol$\,=\,0.7 (red, model 4 in Table~\ref{tbl:nonflatlcdm}), and $\om$\,=\,0.9 and $\ol$\,=\,0.1 (blue, model 6 in Table~\ref{tbl:nonflatlcdm}). The slope and intercept are assumed to evolve with redshift as $\gamma(z)=\gamma_0+\gamma_1(1+z)$ and $\beta(z)=\beta_0+\beta_1(1+z)$. Contours are at 68\% and 95\% confidence levels. The input values of all the correlation parameters are marked by the black square. See Section~\ref{subsect:spaciallyflatlCDMmodel} for details.}
\label{fig:simcheck-evo}
\end{figure*}

\begin{figure*}[h!]
\centering
\includegraphics[width=\linewidth,clip]{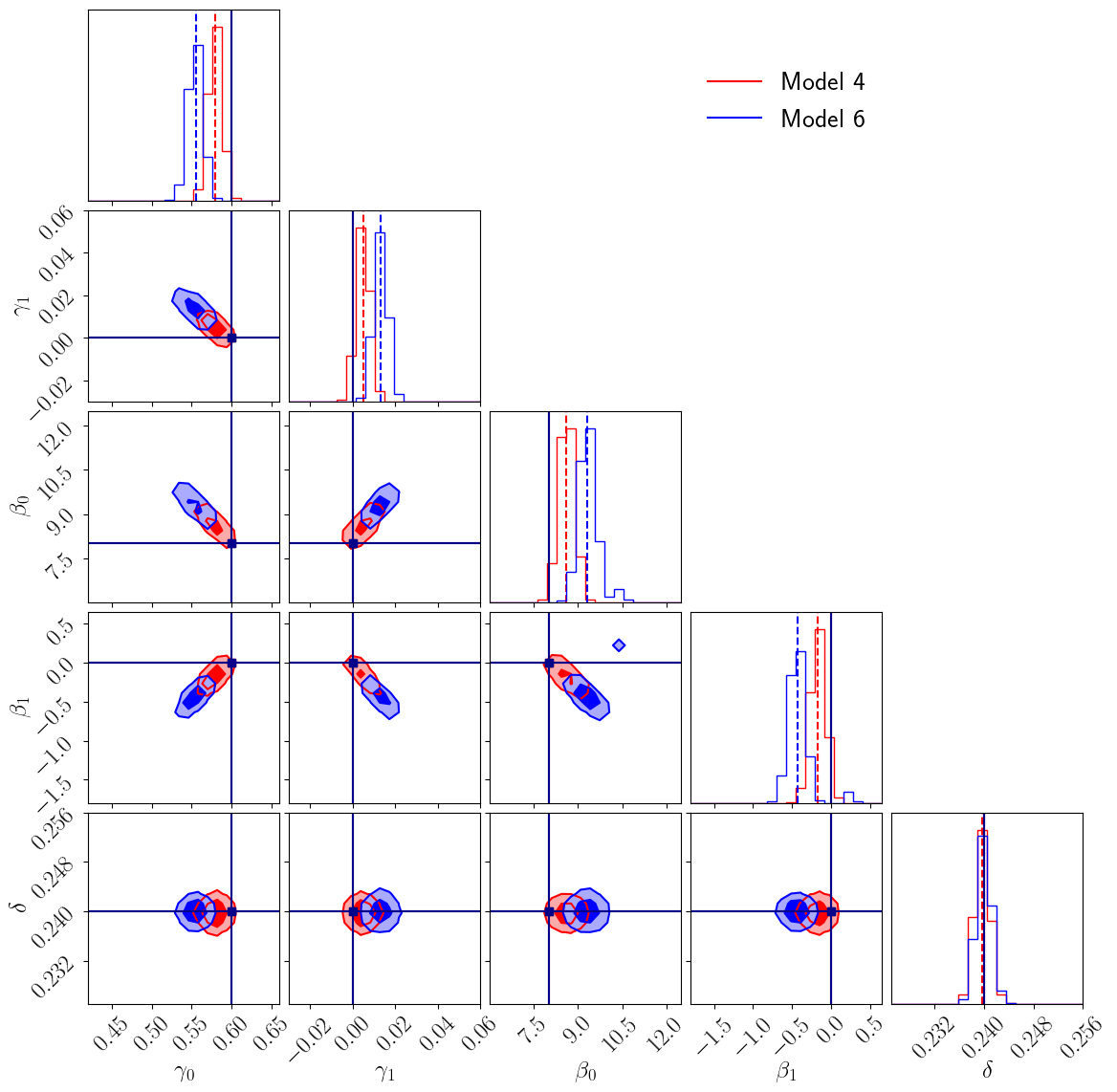}
\caption{Same as in Figure~\ref{fig:simcheck-evo} for sample 2.}
\label{fig:simcheck-evo-big}
\end{figure*}

We finally fit samples 1 and 2 by allowing also for an evolution with redshift of both slope and intercept of the X-ray to UV relation, of the form: $\gamma(z)=\gamma_0+\gamma_1(1+z)$ and $\beta(z)=\beta_0+\beta_1(1+z)$. Figures~\ref{fig:simcheck-evo} and \ref{fig:simcheck-evo-big} show the results of this test. For sample 1, all the correlation parameters are in broad statistical agreement with the true values irrespective of the assumed cosmological parameters, which are fixed to the values $\om$\,=\,0.3 and $\ol$\,=\,0.7 (model 4) or $\om$\,=\,0.9 and $\ol$\,=\,0.1 (model 6). 
The evolution parameters $\gamma_1$ and $\beta_1$, in fact, are consistent with zero within $\sim$\,1.5$\sigma$ at most. 
However, when we appreciably increase the sample statistics, the farther the cosmological model from the true values, the more 
significant the emergence of an evolution of the correlation parameters (Table~\ref{tbl:nonflatlcdm-big}, Figure~\ref{fig:simcheck-evo-big}), even though the input mock sample was built without assuming any evolution. 
This effect is clearly due to the fact that the correlation parameters try to compensate for the deviation of the cosmological model at issue from the input one we had assumed as `true'.


Let us now suppose that the input cosmological model considered to build the mock quasar sample is unknown, and that the adopted working model is very different from the `true' one; any inconsistency between the data and the current model may lead to the wrong conclusion that there is a standardisability problem with the data, rather than a defect in the model that precludes a good representation of the data. 
This is especially true when a potential evolution of the correlation parameters is taken into account: it is not possible to distinguish between the non-standardisability of the data and a limitation of the model itself. 

All these tests support the argument that cosmological models cannot be employed to draw any conclusion on the validity of any given observed data set for cosmological purposes. Along these lines, the now widely-accepted accelerated expansion of the Universe could have been seen as a standardisability issue with supernovae. In the case of blue quasars, a mere comparison of the values obtained for the correlation parameters, $\gamma$ and $\beta$, by fitting any quasar sample (or subsample) within different cosmological models is intrinsically flawed. The results of such an experiment do not (and cannot) entail any cosmology dependence of $\gamma$ and $\beta$.

\section{On the evolution of X-ray and UV luminosities}
\label{sect:On the intrinsic luminosity evolution}
Some authors \citep[e.g.][]{dainotti2022,wang2022,wang2024} have applied a correction for the evolution of both X-ray and UV luminosities with redshift, with the aim of retrieving the `intrinsic', de-evolved correlation between the two luminosities. For instance, by analysing a sample of AGN from SDSS-DR7 (with an $i$-band magnitude $\leq$19.1) cross-matched with {\it Chandra} and {\it XMM-Newton}, \cite{singal2022} determine that the best-fit intrinsic power-law correlation between the X-ray and UV luminosities should be much flatter than what has ever been observed in any AGN sample, i.e., $\gamma=0.28\pm0.03$. 
This value results from the application of the Efron--Petrosian non-parametric method \citep{ep1992}, where the evolution of the luminosities with redshift is assumed in \citet[][see their Section 3.2 for details]{singal2022} to have the following form: 
\begin{equation}
    L^\prime = L/g(z),
\end{equation}
where the evolution function $g(z)$ is defined as:
\begin{equation}
    g(z)=\frac{(1+z)^\eta}{1+\left( \frac{1+z}{Z_{\rm cr}}\right)^\eta},
\end{equation}
with $Z_{\rm cr}=3.7$, while $\eta$ takes the values of 3.3 and 0.55 for the optical and the X-ray data, respectively. 

In practical terms, this correction is largely alternative to allowing for a redshift-evolution of $\beta$. The results are strongly dependent upon both the function and the values of the exponents assumed. Indeed, by applying the same technique but with a simpler evolution function, specifically $g(z)=(1+z)^\eta$, with the exponent taking the values of 4.36 and 3.36 for the optical and the X-ray data, respectively, \citet[][see also \citealt{dainotti2022} and references therein]{Lenart2023} found that the intrinsic slope of the X-ray to UV luminosity relation is statistically consistent with 0.6 instead. The Efron--Petrosian test statistics applies to heavily truncated data, as those delivered by blind surveys. Neither the UV data from the SDSS, nor the X-ray ones after the Eddington-bias correction (see Section~\ref{Eddington bias}) can be plainly considered as such. \citet[][]{Bryant2021} pointed out that the Efron--Petrosian test also depends in a critical way on the choice of the detection threshold. An inaccurate implementation of this method can lead to unreliable or biased conclusions, specifically favouring a stronger, when not spurious, redshift evolution. For these reasons, we believe that this kind of analysis, at this stage, is not a compelling argument against the standardisability of quasars.

\section{General guidelines on how to use quasars for cosmology}
\label{sect:General guidelines}
In this section we outline some general guidelines that should be followed to safely use quasars for cosmology in the framework of the X-ray to UV correlation, through the application of equations (\ref{fxfo}) and (\ref{betahat}). 

\subsection{Sample selection} 
To build a quasar sample that can be utilised for cosmological purposes, multiwavelength data from optical to X-rays are required to derive the rest-frame 2-keV and 2500-\AA\ fluxes. Radio-bright quasars and broad-absorption-line (BAL) sources should be removed to reliably compute X-ray and UV nuclear fluxes that are the direct output of the accretion process and are not affected by jets and/or outflows. Further care should be taken to select low-redshift quasars with minimal contribution from the host galaxy (otherwise a redshift cut at $z$\,$\sim$\,0.7 is a conservative, yet reasonable choice), and sources with negligible flux attenuation from gas and dust at both X-ray and UV energies. Starting from any parent sample, the main aim is to pinpoint the intrinsic, nuclear emission from the accretion disc and the X-ray corona, thus taking advantage of the universal nature of the black-hole accretion process.

Even so, the large dispersion characterising the Hubble diagram of quasars ($\sim$\,1.4 dex) compared to that of type Ia supernovae ($<$\,0.1 dex) still represents a major limitation for the use of quasars as cosmological probes. 
To overcome this large dispersion, some authors suggest to select for cosmological fits quasar subsamples that have a smaller scatter along the X-ray to UV relation, through the application of severe sigma-clipping (down to a threshold of 1.5; \citealt{dainotti2024,dainotti2024b}). However, this procedure is not based on any physical argument, as the parent sample should already be the outcome of a homogeneous selection. Moreover, also the statistical grounds are questionable, since the choice of a tight sigma-clipping threshold has the only effect of cutting the wings of both the X-ray and UV flux distributions. We believe instead that the best way forward is to reduce the overall dispersion \citep[e.g.,][]{signorini2024}, which requires more effort to better understand the physics behind the X-ray to UV relation and to improve the data quality of the observations. 

\subsection{Correction for the Eddington bias} 
\label{Eddington bias} 
Any flux limited sample is biased towards brighter sources at high redshifts, and this effect is more relevant to the X-rays since the relative observed flux interval is narrower than in the UV. To minimise this bias, one possible approach is to neglect all the X-ray detections below a certain threshold, defined as $\kappa$ times the intrinsic dispersion of the $F_{\rm X}-F_{\rm UV}$ relation ($\delta$) computed in narrow redshift intervals (\citealt{lr16,rl19}). Specifically:
\begin{equation}
\label{fthr}
\log F_{2\,\rm keV,\,exp} - \log F_{2\,\rm keV,\,min} < \kappa \delta,
\end{equation}
where $F_{2\,\rm keV,\,min}$ represents the flux limit of a given observation or survey at the rest-frame 2 keV, whilst the product $\kappa \delta$ takes a value that should be estimated for the considered quasar sample. The parameter $F_{2\,\rm keV,\,exp}$ is the monochromatic flux at 2 keV expected from the observed rest-frame quasar flux at 2500 \AA, with the assumption of an {\it intrinsic} value for $\gamma$ of 0.6, and it is calculated as follows:
\begin{equation}
\label{fexp}
\log F_{2\,\rm keV,\,exp} =(\gamma-1)\log(4\pi) + (2\gamma-2)\log D_L + \gamma\log F_{\rm UV} + \beta,
\end{equation}
where $D_L$ is the luminosity distance calculated for each redshift with a fixed cosmology (the results do not depend upon the choice of any reasonable model), and the parameter $\beta$ represents the pivot point of the non-linear relation in luminosities, $\beta=26.5-30.5\gamma\simeq8.0$. 
The Eddington bias can be then reduced by including only X-ray detections for which the minimum detectable flux $F_{2\,\rm keV,\,min}$ in a given observation is lower than the expected X-ray flux $F_{2\,\rm keV,\,exp}$ by a factor proportional to the dispersion in the $F_{\rm X}-F_{\rm UV}$ relation in narrow redshift bins (see Appendix~A in \citealt{lr16} and \citealt{rl19}).

\subsection{Cosmological analysis: the likelihood} 
\label{Cosmological analysis: the likelihood} 
Fitting fluxes with a Gaussian prior on both $\gamma$ and $\beta$ should be the best option under the assumption that the intrinsic values for the slope and intercept do not vary, i.e., the $\lx-\lo$ is universal and redshift-independent (see Section~\ref{sect:gammaevolution}). 
An additional parameter should always be considered to perform the cross-calibration with supernovae, as in equation (\ref{dm}). Nonetheless, it is good practice to fit the quasar sample with variable correlation parameters to verify that their values are in agreement with the assumed ones (i.e., $\gamma$\,$\simeq$\,0.6 and $\beta$\,$\simeq$\,8.0). 
Moreover, by fitting the data with a variable slope, the resulting uncertainties on the output parameters are more conservative than those obtained in the case of fixed correlation parameters.
The likelihood, when fluxes are fitted, can be written as follows:
\begin{equation}
\label{likelihood-fl}
    \ln \mathcal{L}=-\frac{1}{2} \sum_{i=1}^{n}\left[ \frac{(\log F_{{\rm X},i}^{\rm obs}-\log F_{{\rm X},i}^{\rm mod})^2}{\left(d\log F_{{\rm X},i}^{\rm obs}\right)^2+e^{2\ln\delta}} + \ln \left[\left(d\log F_{{\rm X},i}^{\rm obs}\right)^2+e^{2\ln\delta}\right]\right],
\end{equation}
where: 
\begin{multline}
    \log F_{{\rm X},i}^{\rm mod}=\beta+\gamma(\log F_{{\rm UV},i}^{\rm obs}+27.5)-2(1-\gamma)({\rm DM}-28.5)+\\+(\gamma-1)\log(4\pi),
\end{multline}
DM\,=\,DM$(\om,\ol,z,K)$ as given in equation (\ref{dm}), and all the data are in cgs units.

Another possibility is to directly fit the distance moduli, in which case the likelihood in equation (\ref{likelihood-fl}) is modified as follows:
\begin{equation}
\label{likelihood-dl}
    \ln \mathcal{L}=-\frac{1}{2} \sum_{i=1}^{n}\left[ \frac{({\rm DM}_{i}^{\rm obs}-{\rm DM}_{i}^{\rm mod})^2}{\left(d {\rm DM}_{i}^{\rm obs}\right)^2+e^{2\ln\delta}} + \ln \left[\left(d {\rm DM}_{i}^{\rm obs}\right)^2+e^{2\ln\delta}\right]\right],
\end{equation}
where the luminosity distance is:
\begin{equation}
    D_L = 5\left\{\frac{1}{2(1-\gamma)} \left[\beta+\gamma(\log\fo+27.5)-\log\fx\right]+28.1\right\},
\end{equation}
in units of cm.
To quantify the difference between these two approaches, we fit the L20 sample together with the Pantheon+ sample \citep{Scolnic2022}
in a spacially flat $\Lambda$CDM with $\om$ and $\ol$ free to vary. 
The cosmological fit of quasars and supernovae considering fluxes has $\gamma$ and $\beta$ free to vary, whilst the cosmological fit considering luminosity distances assumed fixed values for $\gamma$ and $\beta$ (i.e., $\gamma=0.61\pm0.02$ and $\beta=-31.16\pm0.01$, where the uncertainties are considered for the error propagation on the distances). As $\beta$ is fixed, a cross-calibration parameter $K$ is included in both cases to allow for extra flexibility. As expected, the value of $K$ is very close to zero, implying that the input value for $\beta$ is almost correct.

\begin{figure}[h!]
\centering
\includegraphics[width=\linewidth,clip]{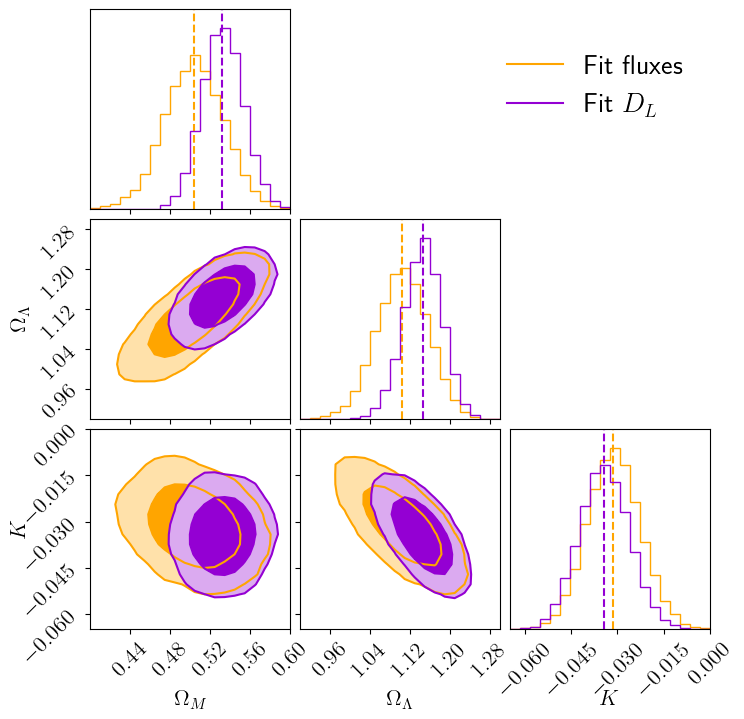}
\caption{Results of the cosmological fit for the L20 quasar sample combined with Pantheon$+$ supernovae in a spacially flat $\Lambda$CDM with $\om$ and $\ol$ free to vary. Orange contours: cosmological fit of quasars and supernovae considering fluxes, with $\gamma$ and $\beta$ free to vary. Violet contours: cosmological fit of quasars and supernovae considering luminosity distances, with $\gamma$ and $\beta$ fixed (i.e., $\gamma$\,=\,0.61 and $\beta$\,=\,$-31.16$). A cross-calibration parameter $K$ is also considered in both cases.}
\label{fig:datafitcheck}
\end{figure}

In Figure~\ref{fig:datafitcheck} we present the comparison of the cosmological parameters, $\om$, $\ol$, and $K$, which clearly shows that they are in statistical agreement within $1\sigma$, with slightly larger uncertainties for the case of variable $\gamma$ and $\beta$, as expected. The summary of the marginalized one-dimensional best-fitting parameters with $1\sigma$ confidence intervals is provided in Table~\ref{tbl:dl_fl}.

\begin{table*}[h!]
\caption{\label{tbl:dl_fl} Marginalized one-dimensional best-fitting parameters with $1\sigma$ confidence intervals for the L20 quasar sample combined with the Pantheon$+$ sample. }
\centering
\begin{tabular}{lccccccc}
\hline\hline
& $\om$ & $\ol$ & $K$ & $\gamma$ & $\beta$ & $\delta$ \\
\hline
Fluxes (free $\gamma$ and $\beta$) & $0.50\pm0.03$ & $1.10\pm0.05$ & $-0.031\pm0.009$ & $0.611\pm0.01$ & $-31.46\pm0.01$ & $0.23\pm0.01$ \\
DM (fixed $\gamma$ and $\beta$) & $0.53\pm0.02$ & $1.15\pm0.04$ & $-0.034\pm0.008$ & - & - & $1.44\pm0.03$ \\
\hline
\end{tabular}
\tablefoot{The cosmological model considered is a spacially flat $\Lambda$CDM with $\om$ and $\ol$ free to vary. We compare the results of two different fitting procedures: fitting X-ray and UV fluxes with $\gamma$ and $\beta$ free to vary, and fitting the distance moduli with fixed $\gamma$ and $\beta$ values (i.e. $\gamma$\,=\,0.61 and $\beta$\,=\,$-31.16$). A cross-calibration parameter $K$ is also considered in both cases.}
\end{table*}

\section{Conclusions}
\label{sect:conclusion}
The aim of this work is to address the question whether the non-linear relation between the X-ray and UV emission of quasars can be employed to derive quasar distances and, therefore, typical blue quasars can be regarded as standardisable candles for cosmology. For such a technique to be reliable, the correlation parameters $\gamma$ (slope, in log space) and $\beta$ (intercept) must be cosmology- and redshift-independent. As already shown in previous works of our group, the distances estimated with this method agree with the standard flat $\Lambda$CDM model up to $z$\,$\sim$\,1.5, but they start to deviate significantly from the model predictions at higher redshifts. For this reason, some authors suggested that the observed discrepancy could be due to heterogeneities between the low- and high-redshift subsamples considered in \citet{lusso2020}, or to an evolution of the relation with redshift, ultimately implying the non-standardisability of quasars. To refute the latter arguments, here we argue that the values of $\gamma$\,$\simeq$\,0.6 and $\beta$\,$\simeq$\,8.0 are inherent to the black-hole accretion process and consequential to its universal nature, thus being intrinsically cosmology-independent. Indeed, we have demonstrated that the relation between X-ray to UV fluxes does not show any redshift evolution in its slope when studied in narrow redshift bins, confirming that $\gamma$ is both redshift- and cosmology-independent. Moreover, we performed a set of simulations showing that all the claimed inconsistencies naturally arise from any limitations of the cosmological model(s) used in the analysis, and do not necessarily indicate any problem with the data set itself. Specifically, the claimed non-standardisability of high-redshift quasars is an obvious manifestation of our ignorance of the true cosmological model. 
Our main results can be summarised as follows:
\begin{itemize}
    \item The fit the $\fx-\fo$ relation in narrow redshift bins and the comparison with supernova-derived distances in the common redshift range, show that quasars are fully consistent with supernovae and that there is no redshift evolution of the slope $\gamma$ at $z$\,$<$\,1.6. The statistical analysis of the quasar sample is then extended up to $z$\,$\simeq$\,3.5, confirming that the $\gamma$ parameter does not exhibit any systematic trend in the entire redshift range over which we have enough statistics to perform the regression fit. This result does not depend on the width of the adopted redshift bins, provided dispersion in luminosity distance within each bin is smaller than the one expected for the X-ray to UV relation. The deviation from the flat $\Lambda$CDM we start to detect beyond $z$\,$\approx$\,1.6 cannot thus be attributed to an evolution of the slope $\gamma$. While no direct test can be performed to also verify the non-evolution of $\beta$, its sudden change at $z$\,$>$\,1.6 is not supported by observational evidence.
    \item Simulations of a mock quasar sample demonstrate that, when different cosmological models are compared, there is an unavoidable degeneracy between the detection of a possible evolution of $\gamma$ and $\beta$ (or a departure from their `intrinsic' values), and the determination of the cosmological parameters. 
    This is a general feature of the standard-candle method: any departure from the correct, unknown cosmology can be neutralised through a suitable modification or evolution of some parameter intrinsic to the physical sources in use, whether these are quasars or supernovae. Hence, no cosmological models or comparison thereof can conclusively validate or reject any given probe, or data set. 
\end{itemize}

We conclude that the reliability of our method is fully corroborated by the comparison with supernovae up to $z$\,$\simeq$\,1.5, while outside the common redshift range it can only derive from a cosmology-independent evaluation of the hypothesis of non-evolution, especially for $\beta$, subject to a meticulous check of the sample selection and of the flux measurements for possible redshift-dependent systematic effects. Multi-wavelength observations of quasars across a wide redshift range show a high level of homogeneity in their spectral properties. As a matter of fact, when a supermassive black hole steadily accretes matter, the physics underlying this process is the same regardless of redshift; and so is its observational appearance in the X-ray and UV bands, as long as no additional energy-extraction mechanism (e.g., the launch of radio jets and/or fast uncollimated outflows) or contamination effects (e.g., host-galaxy emission, dust/gas absorption are present.
Since there are no clear systematics in the sample selection, nor in the flux measurements for all the quasars we have employed in our works, we are confident the application of the X-ray to UV relation to cosmology is solid and well motivated. 
A better understanding of the physical process behind the $\lx-\lo$ relation can definitely strengthen the method, and more effort is required along this line. Future measurements of supernovae at redshifts $z$\,$>$\,2 possibly confirming the discrepancy with $\Lambda$CDM found with quasars, will provide an independent observational proof that new physics is involved.

\begin{acknowledgements}
We thank the anonymous reviewer for their thorough reading and for useful comments. This work was performed in part at Aspen Center for Physics, which is supported by National Science Foundation grant PHY-2210452. All the authors acknowledge partial support from the Simons Foundation (1161654, Troyer).
\end{acknowledgements}

%
   \bibliographystyle{aa} 
   \bibliography{bibl} 
%
\end{document}